\PassOptionsToPackage{dvipsnames}{xcolor}
\PassOptionsToPackage{hyphens}{url}

\documentclass[10pt, conference, letterpaper, anonymous]{IEEEtran}

\usepackage{cite}
\usepackage{todonotes}
\usepackage{amsfonts}  % Mathematical sets
\usepackage{gensymb}   % Symbols
\usepackage{algpseudocode}
\usepackage{algorithm}
\usepackage{forest}  % Draw trees in TikZ
\usepackage{xcolor}
\usepackage{pifont}  % Checkmarks and crosses
\usepackage{subcaption}  % Subfigures
\usepackage{multirow}  % Tables: cells spanning multiple rows
\usepackage{makecell}
\usepackage{listings}  % Code listings
\usepackage{hyperref}
\usepackage{stackengine}  % Overlapping images
\usepackage[normalem]{ulem}  % Strikethrough text
\def\BibTeX{{\rm B\kern-.05em{\sc i\kern-.025em b}\kern-.08em
    T\kern-.1667em\lower.7ex\hbox{E}\kern-.125emX}}

\newcommand*\circled[1]{\raisebox{.5pt}{\textcircled{\raisebox{-.9pt} {#1}}}}
\newcommand*\step[1]{\textcolor{orange}{\circled{#1}}}

\newcommand{\xmark}{\textcolor{Red}{\ding{55}}}
\newcommand{\texturl}[1]{{\small \texttt{#1}}}

\newcommand\rurl[1]{%
  \href{https://#1}{\nolinkurl{#1}}%
}

\usetikzlibrary{automata,positioning,backgrounds,patterns}
\tikzstyle{lan}    = [draw=blue, semithick]
\tikzstyle{wlan}   = [draw=blue, dotted, thick]
\tikzstyle{wan}    = [draw=Green, semithick]
\tikzstyle{zigbee} = [draw=red, dotted, thick]
\tikzset{
  every node/.style={font=\footnotesize},
  connector/.style={
      -stealth,
      font=\scriptsize
  },
  rectangle connector/.style={
      connector,
      to path={(\tikztostart) -- ++(#1,0pt) \tikztonodes |- (\tikztotarget) },
      pos=0.5
  }
}

\AtBeginDocument{%
  \providecommand\BibTeX{{%
    Bib\TeX}}}

\IEEEoverridecommandlockouts\IEEEpubid{\makebox[\columnwidth]{979-8-3315-0376-5/25/\$31.00 $\copyright$2025 IEEE \hfill}\hspace{\columnsep}\makebox[\columnwidth]{ }}

\begin{document}

\title{The Forest Behind the Tree: Revealing Hidden Smart Home Communication Patterns}

%% AUTHORS %%

\makeatletter
\newcommand{\linebreakand}{%
  \end{@IEEEauthorhalign}
  \hfill\mbox{}\par
  \mbox{}\hfill\begin{@IEEEauthorhalign}
}
\makeatother

%% Uncomment for anonymous submission
%\author{\IEEEauthorblockN{Anonymous Author(s)}}

%% Uncomment for camera-ready
\author{
  \IEEEauthorblockN{François De Keersmaeker\textsuperscript{1}, Rémi Van Boxem\textsuperscript{2}, Cristel Pelsser\textsuperscript{1}, Ramin Sadre\textsuperscript{1}}
  \IEEEauthorblockA{
  \textsuperscript{1}UCLouvain, Louvain-la-Neuve, Belgium - \{francois.dekeersmaeker, cristel.pelsser, ramin.sadre\}@uclouvain.be \\
  \textsuperscript{2}Junia, Lille, France - remi.van-boxem@student.junia.com
  }
% \and
%   \IEEEauthorblockN{Rémi Van Boxem}
%   \IEEEauthorblockA{
%   \textit{Junia}\\
%   Lille, France \\
%   remi.van-boxem@student.junia.com}
% \and
%   \IEEEauthorblockN{Cristel Pelsser, Ramin Sadre}
%   \IEEEauthorblockA{
%   \textit{UCLouvain}\\
%   Louvain-la-Neuve, Belgium \\
%   \textit{firstname}.\textit{lastname}@uclouvain.be}
}


\maketitle

\begin{abstract}

The widespread use of Smart Home devices has attracted significant research interest in understanding their behavior within home networks. Unlike general-purpose computers, these devices exhibit relatively simple and predictable network activity patterns. However, previous studies have primarily focused on normal network conditions, overlooking potential hidden patterns that emerge under challenging conditions. Discovering the latter is crucial for assessing device robustness.

This paper addresses this gap by presenting a framework that systematically and automatically reveals these hidden communication patterns. By actively disturbing communication and blocking observed traffic, the framework generates comprehensive profiles structured as behavior trees, uncovering traffic flows that are missed by more shallow methods. This approach was applied to ten real-world devices, identifying 254 unique flows, with over 27\% only discovered through this new method. These insights enhance our understanding of device robustness, and the thus obtained profiles provide a more complete description of the network behavior of devices, as needed, for example, for the configuration of security solutions.
\end{abstract}


%% BODY TEXT SECTION %%
\section{Introduction}\label{sec:intro}

In the \textbf{I}nternet \textbf{o}f \textbf{T}hings (IoT) paradigm, physical objects are equipped with sensing, computing, and networking capabilities. This allows them to monitor or react to their environment and exchange messages with other objects and with remote servers \cite{iot_survey}.
A common use case of this paradigm is the Smart Home, composed of household objects, ranging from small power outlets to large appliances. The popularity of Smart Home devices has increased sharply in the past years, with a market size estimated at 101.07 billion USD in 2024 \cite{smart-home-market-share-fortune}.
Their main appeal is to provide home automation, and therefore convenience, to the user.

As IoT devices have very low computing resources,
it is a technical challenge to embark them with all the technology
necessary for their correct operation.
Notoriously, to stay economically competitive,
manufacturers tend to expand the user-visible functionalities
on the devices' limited hardware,
neglecting transversal aspects such as robustness or security \cite{toys,in-the-room}.
Ironically, this compromise might impede the initial incentive behind the design of Smart Home systems,
i.e. user convenience.
Indeed, if devices do not provide robustness in the face of network communication
instability, they may become unresponsive,
disrupting the whole system.

The motivation behind this work is to assess the robustness of Smart Home devices
against network instabilities.
More precisely, we aim to discover and describe network communication patterns issued by such devices
in the case where their default traffic fails.

% For these reasons, firewalls and network-based \textbf{I}ntrusion \textbf{D}etection \textbf{S}ystems (IDS) for Smart Homes have been extensively researched in the past decade \cite{passban_ids, argus, duan_iota_2023, smart-home-firewall}. The aim of such security solutions is to detect and block unwanted network communications from and to the devices. Typically, the firewall or IDS is given a description of the traffic allowed and/or disallowed for a particular device. This description, called a \emph{device profile} in the following, must be as complete as possible, as an incomplete profile would cause the firewall or IDS to block legitimate or allow unwanted communication attempts. Unfortunately, only very few manufacturers provide insight into the communication behavior of their IoT devices and services, although the IETF has even standardized a format dubbed the Manufacturer Usage Description (MUD) \cite{mud} to write down profiles in a compact, machine-readable form, albeit admittedly relatively limited \cite{matheu_extending_mud_2019, singh_clearer_than_mud_2019, smart-home-firewall}.

%As they are often programmed for a specific task,
Smart Home devices usually exhibit simple and predictable
network communication patterns \cite{sivanathan_classifying_2019} under ideal conditions.
Consequently, researchers have designed solutions to express their network behavior
in a compact form, a so-called \emph{profile}.
The IETF has even standardized a format for such profiles,
the \textbf{M}anufacturer \textbf{U}sage \textbf{D}escription (MUD) \cite{mud},
albeit admittedly relatively limited \cite{matheu_extending_mud_2019, singh_clearer_than_mud_2019, smart-home-firewall}.
Several methods have been presented in the past to automatically create the profile of a device \cite{hamza_clear_2018,ping-pong,homesnitch}. For this purpose, its network communications are observed over a certain period of time, in the wild or under laboratory conditions. Typically, relatively complex experimental setups are required \cite{saidi_haystack_2020, behaviot}, due to the variety of possible interactions with the device.
Data mining and analysis techniques are then used to extract compact representations of the network traffic and associate them with the device,
or even with specific device events.

%In order to gain a comprehensive view on the communication profile of a device, relatively complex experimental setups are required \cite{saidi_haystack_2020, behaviot}, due to the variety of possible interactions with the device.
%, ranging from simply using the manufacturer-provided smartphone application,
%to leveraging a third-party \emph{Smart Home automation platform}
%where the device is part of a user-defined routine.

%
% RAMIN: I don't think we should talk about the quality of signature extraction algorithms,
% since this is not a topic we address in the paper.
%
%% Signature extraction algorithms
%However, existing algorithms to extract signatures for device events from network traffic
%are quite complex.
%Numerous leverage demanding \textbf{M}achine \textbf{L}earning (ML) models,
%requiring a heavy quantity of labeled data,
%which can hinder their applicability in a wide range of systems.
%Moreover, whereas such algorithms indeed usually showcase very accurate results
%when applied to the dataset and system they were trained on,
%their usefulness tends to decline when applied to different systems \CP{Add ref}.
%We argue that the task at hand, i.e. extracting event signatures from packet captures,
%can be achieved with a simpler algorithm,
%i.e. merely extracting the network flows which have occurred
%in all network captures corresponding to successful event executions.

We argue that existing approaches to profiling are insufficient for our goal of gaining comprehensive insight into the network behavior of Smart Home devices, as they do not aim at triggering communication patterns that only become visible under certain network conditions, in particular when the default communication mechanism does not succeed.
For example, an IoT device might have a list of alternative domain names for its cloud hosted services that it will only contact if the default domain name fails to resolve or the server behind that name does not respond;
or it can switch from using TCP to UDP (and vice-versa), if one of the protocols is blocked.
Ignoring such behaviors leads to profiles, in which some of the device's communication patterns are missing or incompletely described, making it difficult to grasp if a device is robust with regard to network events.

In this paper, we present a framework, mostly automated and requiring little prior configuration, to comprehensively uncover the communication patterns of a Smart Home device. For each type of interaction with the device, it first observes the traffic flows that occur in an unconstrained network and builds a set of flow descriptors from them.
Such descriptors contain Internet and transport-layer information,
as well as selected, available application-layer data,
such as domain names.
The system then iteratively blocks flows corresponding to certain descriptors and repeats the interaction with the intent of making new flows appear. 
The algorithm selecting which flows to block and the sequence of blocking experiments is not straightforward.
Our aim is to discover as many hidden flows as possible but keep the number of experiments under control. 
To do this efficiently and not end up in an infinite loop where the same patterns appear over and over again, we store the found descriptors in a tree that we traverse using a breadth-first strategy,
and prune using an \textit{ad hoc} heuristic.
The result is a set of \emph{multi-level}, tree-shaped profiles (one for each type of interaction) for the device that describe the device's default and alternative communication patterns that we then explore to assess the robustness of the device's network communications.
%Our profiles are conceptually similar to the aforementioned ones, e.g. MUD,
%and can therefore also be used as input configuration for allow-list firewalls.

Conceptually, the profiles obtained by our approach can be treated as enhanced versions of the aforementioned ones,
such as MUD,
as they encompass the base communication patterns,
and augment them with the identified hidden patterns.
Therefore, they can also be used as input configuration for allow-list firewalls.
Current state-of-the-art approaches,
which do not cover alternative communication patterns,
might consider the latter as malicious,
whereas they are actually part of the device's intended behavior,
only rarer. 
Profiles obtained from our framework can thus provide more accurate security configurations that do not block the legitimate communications of a robust device.

%In the context of this work, we then leverage the resulting signatures to assess the robustness
%of the devices' network communications.
%However, they might also be repurposed and be used as configuration
%for allow-list firewalls.

% Workflow overview
%In this work, we position ourselves from the perspective of Smart Home events;
%our objective is then to apply our aforementioned strategy to profile the covered events.
%For each event, we will produce a \emph{multi-level profile},
%comprising the first-level traffic patterns,
%as well as the newly emerged patterns.
%In practice, we instrument a corpus of Smart Home devices,
%issue their typical events (e.g. for a smart plug, toggle it),
%and extract the event's network signature.
%We repeat this,
%while iteratively blocking the previously-seen signatures,
%with the intent of making new signatures appear.
%As we want to trigger all the event's potential signatures,
%we stop iterating when no new signature occurs.

% Contributions
The contributions of our work can be summarized as follows:
\begin{itemize}
%\item \textbf{We design a new, simple algorithm to profile Smart Home devices at the network flow level.}
%For a device interaction, the algorithm builds an event signature, i.e., the description of flows appearing for that interaction.
% by extracting the network flows present at each successful event iteration.
% Our algorithm does not rely on Machine Learning nor heavy statistical computation,
% and therefore does not require a big amount of (labeled) data,
% making it portable and easy to apply in various small networks.

\item \textbf{We develop and implement a framework},
modular and easily extensible,
which instruments Smart Home devices and \textbf{extracts multi-level profiles} for them, in a \textbf{mostly automated way}. The core of the framework is a new algorithm that, for a given device interaction, builds an event signature, i.e., the description of flows appearing for that interaction, and then iteratively blocks previously observed flows, until the interaction fails or no new flow is discovered.

\item \textbf{We apply our framework} to a testbed network comprising ten off-the-shelf devices,
and generate multi-level profiles for 36 unique events,
which sum up to a total of 254 unique network flows discovered,
of which 70 were ``hidden'', i.e., not part of the default communication patterns of the tested devices, accounting for over 27\% of the total number of unique flows, representing the portion that traditional techniques are likely to miss.

\item \textbf{We assess the robustness} of the instrumented devices,
by computing robustness-related metrics.
We conclude that most devices provide at least one backup communication strategy,
if a default communication pattern fails.

\item \textbf{We publish the source code}
of our signature extraction algorithm
at \url{https://github.com/smart-home-network-security/signature-extraction},
and of our experimental framework,
\textbf{including the captures of the testbed's traffic} at
\url{https://github.com/smart-home-network-security/smart-home-hidden-communication-patterns}.
\end{itemize}

% Paper structure
The rest of the paper is structured as follows. We describe the problem and the scope of our work in Section~\ref{sec:problem}.
In Section~\ref{sec:framework},
our methodology is explained in detail.
Its evaluation is presented in Section~\ref{sec:results}.
Section~\ref{sec:related-work} presents related work
and positions them compared to ours.
Ultimately, the paper concludes in Section~\ref{sec:conclusion}.

\section{Problem statement and scope}
\label{sec:problem}

Fig.~\ref{fig:scenario} shows a typical Smart Home setup with a smart lamp that can be controlled by a smartphone application (called \emph{companion app} in the following).
The possible network activities in this setup are highly device specific.
The device might communicate directly with the companion app in the local home network, or do this via a manufacturer or vendor-operated cloud server (the latter making it also possible to control the device from outside the home network and to receive firmware updates). In fact, many of the devices on the market have both local and remote communication capabilities.

\begin{figure}
  \centering
  \begin{tikzpicture}[node distance=2cm,on grid,auto]

    \node (ap) [label={[yshift=-1mm]above:{wireless AP}}] {\includegraphics[scale=0.06] {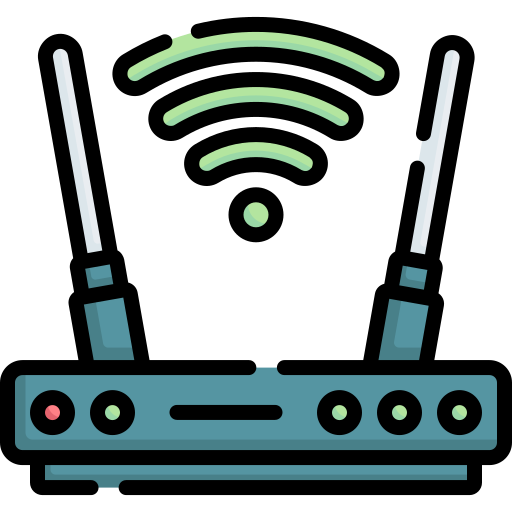}};
    \node (router) [right=1.5cm of ap, label={[align=center, yshift=-2mm]above:{LAN\\gateway}}] {\includegraphics[scale=0.07]{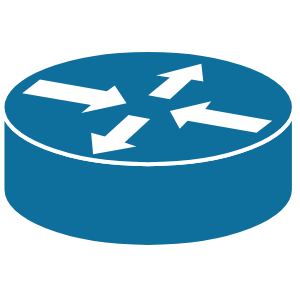}};
    \node (internet) [right=1.7cm of router, label={[yshift=-3.5mm,align=center]above:{vendor-operated\\cloud}}] {\includegraphics[scale=0.08]{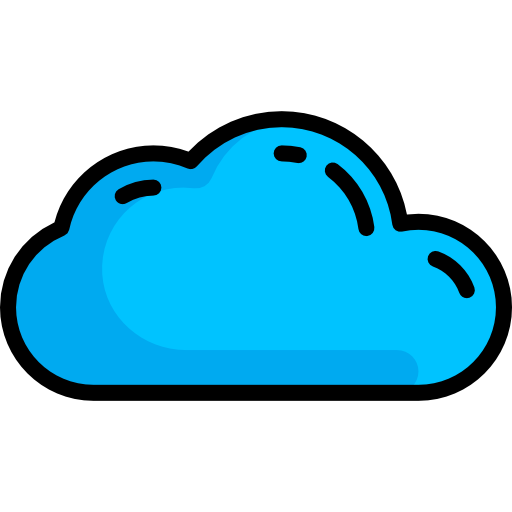}};

    \node (phone) [left=2cm of ap,yshift=0.5cm, label={[align=right, xshift=0.15cm] left:{smartphone\\running\\companion app}}] {\includegraphics[scale=0.05]{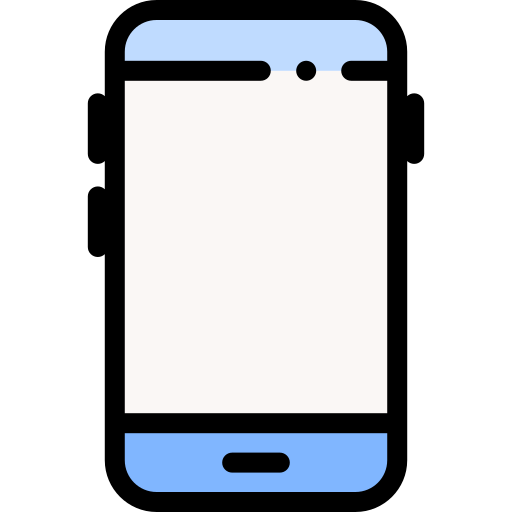}};
    \node (wifi-light)[left=2cm of ap, yshift=-0.5cm, label={[align=right, xshift=0.15cm, yshift=-0.5mm] left:{smart\\lamp}}] {\includegraphics[scale=0.05]{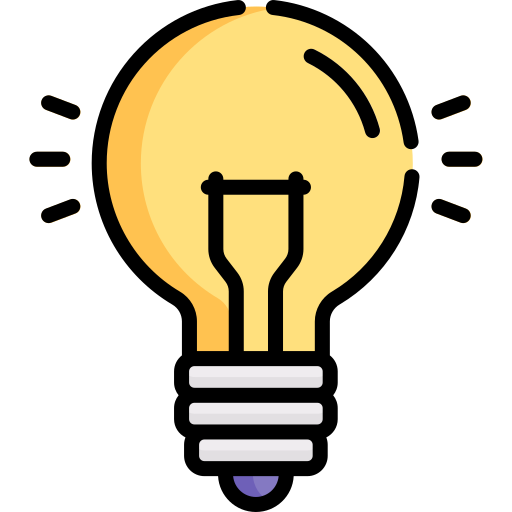}};

    \draw [wlan] (phone)  -- (ap);
    \draw [wlan] (ap) -- (wifi-light);
    \draw [wan] (ap) -- (router);
    \draw [wan] (router)  -- (internet);

  \end{tikzpicture}
  \caption{Typical Smart Home network with a smart lamp}
  \label{fig:scenario}
\end{figure}

In line with existing work on profiling discussed in Section~\ref{sec:related-work}, we assume that the network activity is triggered by an interaction with the Smart Home device. The aim of our work is to obtain a profile, i.e., a description of the network communication patterns associated to that device and interaction. We argue that existing approaches on device profiling, that mostly consist in performing the interaction and recording the resulting network activity, are not sufficient to observe all possible communications patterns.
Indeed, we expect that some of them only appear when the default (or configured) patterns fail or are disturbed. Our goal is to discover such patterns in an automated way.

The solution that we will present in the next sections requires the collection, inspection and filtering of packets sent to and received from the Smart Home device and the device running the companion app. If the traffic is unencrypted, we leverage its payload. We do not analyze or modify the device firmware, the companion app, or the server side software. We also do not try to decrypt network traffic, therefore we do not consider information in encrypted payload. However, numerous research has shown that relying on unencrypted information and packet or flow metadata, can already provide sufficient information to accurately fingerprint IoT devices \cite{homesnitch, ping-pong, wan_iotathena_2022}. 

The terms IoT and Smart Home are quite fuzzily defined in the literature.
Our work focuses on domestic devices,
augmented with IoT capabilities, which exhibit
a precise, goal-oriented, and limited set of functions,
e.g. power plugs, light bulbs, and cameras.
More powerful/general-purpose smart home devices
(e.g. smart TVs or speaker hubs)
do not exhibit such a precise set,
and are therefore considered out of scope.
Speakers usually act as an intermediary between the user and the end device,
enabling the user with voice control.
Additionally, such devices allow installing third-party apps
which provide them with limitless capabilities.
This makes the network behavior of such devices very complex and customizable,
rendering an automated analysis with our framework practically unfeasible.
Indeed, as the set of functions is unlimited and not known \textit{a priori},
a single device can showcase an unbounded amount of different tree profiles,
as one profile pertains to one function.

Similarly to related research works, we only consider IP traffic, either wired or wireless.
Other protocols exist for Smart Home devices, such as Zigbee \cite{zigbee} and Thread \cite{thread}
over IEEE 802.15.4 \cite{802-15-4}, or the proprietary Z-Wave \cite{z-wave}. Such protocols usually rely on a gateway or hub to connect to the local home network. For devices using these protocols, we profile the gateway's IP traffic.
Devices that rely on long-range communication technologies, such as 5G or LoRaWAN \cite{lorawan}, are not in the scope of this paper since their traffic cannot be locally filtered or blocked.

Finally, it should be mentioned that devices also communicate for reasons other than being triggered by interactions. Examples include periodic heartbeat messages or messages to discover the local network \cite{mazhar_characterizing_2020, behaviot}. A possible approach to profile such activities is to passively monitor the device over a longer period and extract the traffic using periodicity models \cite{behaviot}.
Our approach does not pursue such communication activities further, as in this paper we focus on the more complex case of interaction-triggered communication patterns.

% \subsection{Threat model}

% As our work's incentive is to enhance the security of smart home networks,
% we first provide the threat model against which our event profiles are supposed to protect.

% We focus on malicious traffic which deviates from the typical operating behavior of the monitored devices.
% %, either going towards or coming from those.
% The two main scenarios are the following:
% \begin{itemize}
%   \item An attacker, whether inside or outside the local network, who issues traffic towards a device,
% in order to disrupt it (e.g. DoS) or compromise it (e.g. to exfiltrate private user data).
%   \item A compromised device, inside the LAN, issues traffic towards other devices in the LAN (e.g. to compromise other devices by spreading a worm) or towards the Internet (e.g. to communicate with an external server belonging to the attacker).
% \end{itemize}

% In line with related work,
% we assume that, during the "generation" phase of the event profiles,
% our devices are exempt from malicious traffic.
% Once our profiles have been generated and enforced,
% malicious traffic covered by the threat model would be detected and blocked.

\section{Profiling methodology}
\label{sec:framework}

In this section, we describe our profiling methodology. We start with an outline of our approach, followed by a detailed presentation of its components.

\subsection{General workflow}
\label{sec:overview}

Fig.~\ref{fig:workflow} gives an overview on the workflow of our methodology to obtain a profile for a Smart Home device under a certain type of interaction. An interaction can require the usage of a triggering source, such as a companion app installed on a smartphone. In practice, a Smart Home device may support different interactions (e.g., for a smart lamp: switching on/off, changing its color, etc.) and triggering sources, and the workflow described here must be repeated for each of them. The workflow consists of multiple steps summarized below. Besides initial manual configuration in Step 1, the rest of the workflow is automated.

%For all devices, the considered interactions have been chosen to closely mimic a real user interacting with the device in a Smart Home setting, by using its companion app; e.g., for a smart lamp, we consider the following interactions: switching on/off, changing the brightness, changing the color.

\begin{figure}
  \centering
  
  \tikzstyle{flow} = [fill=Cyan, minimum width=1cm, minimum height=9pt, inner sep=2pt]
  \begin{tikzpicture}[
    node distance=2cm,
    on grid,
    auto]

  %%% FIREWALL
  \node (ap) {\includegraphics[scale=0.06]{figures/testbed/wireless-ap.png}};
  \node (firewall)[xshift=0.3cm,yshift=-0.25cm,label={[align=center,yshift=0.2cm]below:{LAN AP\\+ firewall}}] at (ap) {\includegraphics[scale=0.05]{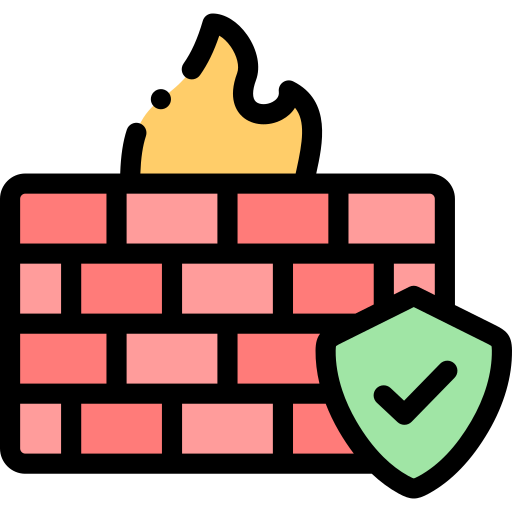}};

  %%% DEVICE EVENTS
  % Device
  \node (device) [left=1.5cm of ap,label={[yshift=0.1cm]below:{device}}] {\includegraphics[scale=0.04]{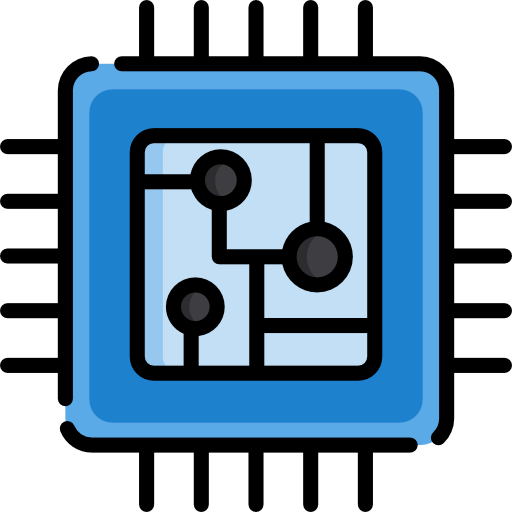}};
  \draw[->,thick,Cyan] (device) -- (ap) node[midway,above,label={[black,yshift=-0.1cm]above:{pkts}}] {\includegraphics[scale=0.02]{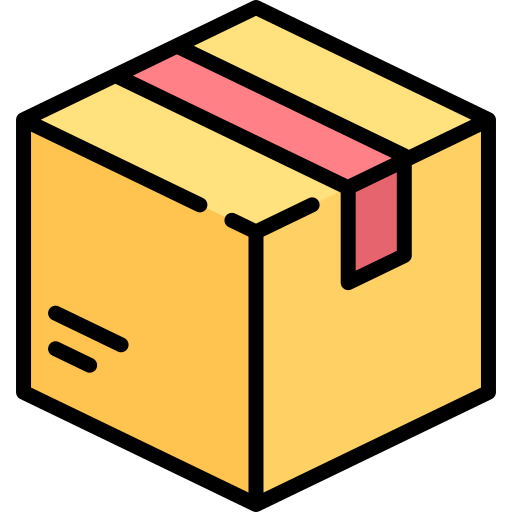}};
  % Event trigger
  \node (event) [left=1.2cm of device,label={[align=center,yshift=0.1cm]below:{interaction\\trigger}}] {\includegraphics[scale=0.05]{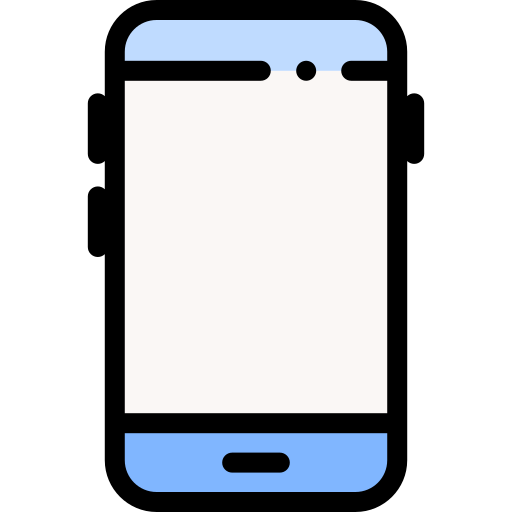}};
  \draw[->,thick] (event) -- (device) node[midway] {\step{1}};

  % PCAP files
  \node (pcap) [right=2cm of ap,label={[yshift=0.1cm]below:{PCAPs}}] {\includegraphics[scale=0.07]{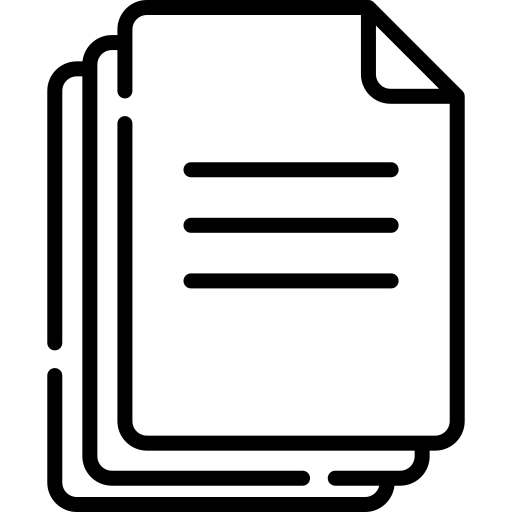}};
  \draw[->,thick] (ap) -- (pcap) node[midway] {\step{2}};

  %%% Signature extraction
  \node (sig-ex) [right=1.7cm of pcap,label={[align=center,yshift=0.1cm] below:{Signature\\extraction}}] {\includegraphics[scale=0.06]{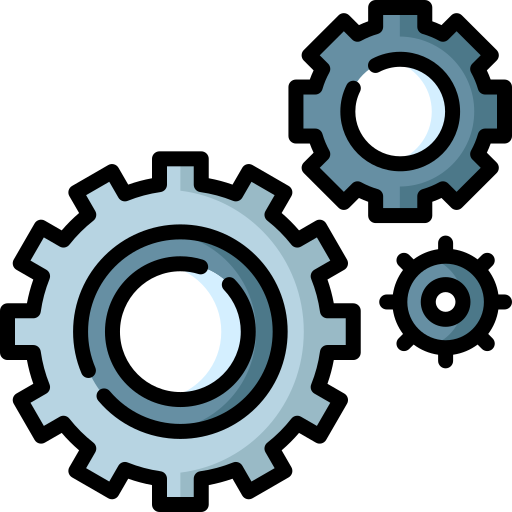}};
  %\node[right=2cm of pcap, rounded corners, fill=blue!20, align=center, text centered] (sig-ex) {Signature\\extraction};
  \draw[->,thick] (pcap) -- (sig-ex) node[midway,xshift=-1mm] {\step{3}};
  % Background
  \begin{scope}[on background layer]
    \coordinate [above=0.6cm of sig-ex,xshift=-0.7cm] (sig-ex-top-left);
    \coordinate [below=1.2cm of sig-ex,xshift=0.7cm] (sig-ex-bottom-right);
    \draw [rounded corners, draw=none, fill=blue!20] (sig-ex-top-left) rectangle (sig-ex-bottom-right);
  \end{scope}

  % Event signature
  \node (ev-sig) [below=2.3cm of sig-ex, label={[yshift=0.1cm,align=center]below:{Event\\signature}}] {\includegraphics[scale=0.07]{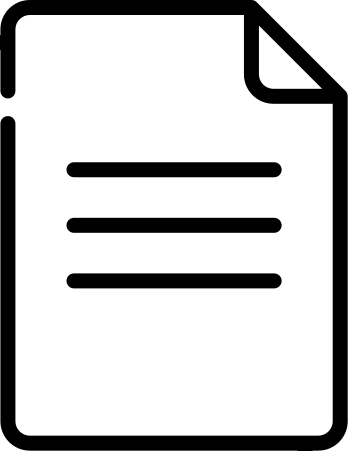}};
  \draw[->,thick] ([yshift=-0.6cm]sig-ex.south) -- (ev-sig) node[midway] {\step{4}};

  % Extracted network flows
  \node[flow, left=1.75cm of ev-sig] (flow-c) {$c$};
  \node[flow, above=0.5cm of flow-c, fill=Cyan] (flow-b) {$b$};
  \node[flow, below=0.5cm of flow-c, fill=Cyan, label={[yshift=-0.2cm]below:{Flow IDs}}] (flow-d) {$d$};
  \coordinate [above=0.3cm of flow-b, xshift=-0.7cm] (rect-top-left);
  \coordinate [below=0.3cm of flow-d, xshift=0.7cm] (rect-bottom-right);
  \draw (rect-top-left) rectangle (rect-bottom-right);

  % Arrows: Event signature -> flows
  \node (split) [circle, fill=black, inner sep=1pt, left=0.75cm of ev-sig] {};
  \draw[thick] (ev-sig) -- (split.center);
  \draw[->,thick] (split.center) -- (flow-c);
  \draw[->,thick] (split.center) |- (flow-b) node[midway,above] {\step{5}};
  \draw[->,thick] (split.center) |- (flow-d);

  %%% TREE
  \begin{scope}[
    font=\scriptsize,
    level distance=0.7cm,
    edge from parent/.style={draw,-latex},
    every node/.style={circle, draw, minimum size=0.5cm, inner sep=0pt},
    level 1/.style={sibling distance=1.6cm},
    level 2/.style={sibling distance=0.6cm}
    ]
    % Draw the binary tree
    \node (tree-root) [left=3cm of flow-c,yshift=1cm] {}
        child {node (tree-left) {}
            child {node {}}
            child {node {}}
        }
        child {node (tree-right) {}
            child {node[dotted] {$b$} edge from parent[dotted]}
            child {node[dotted] {$c$} edge from parent[dotted]}
            child {node[dotted] {$d$} edge from parent[dotted]}
        };

    \draw[red, thick, rotate=-40] (tree-root)[xshift=0.5cm] ellipse (0.9cm and 0.5cm);
  \end{scope}
  \node (tree-caption) [below=2cm of tree-root,align=center] {Multi-level event signature tree};

  % Arrow: flows to tree
  \draw[->,thick] ([xshift=-0.2cm,yshift=-0.4cm]flow-c.west) -- ([xshift=0.8cm,yshift=-0.7cm]tree-right.east) node[midway, above] {\step{6}};

  % Arrow: tree -> AP
  \draw[->, red, thick] ([yshift=0.1cm]tree-root.north) -- (ap) node[midway,below,xshift=1.5mm,yshift=1.2mm] {\step{7}};

  %% Output profile
  \node (profile) [left=1.75cm of tree-left,yshift=0.25cm,label={[align=center,yshift=0.1cm]below:{multi-level\\profile}}] {\includegraphics[scale=0.07]{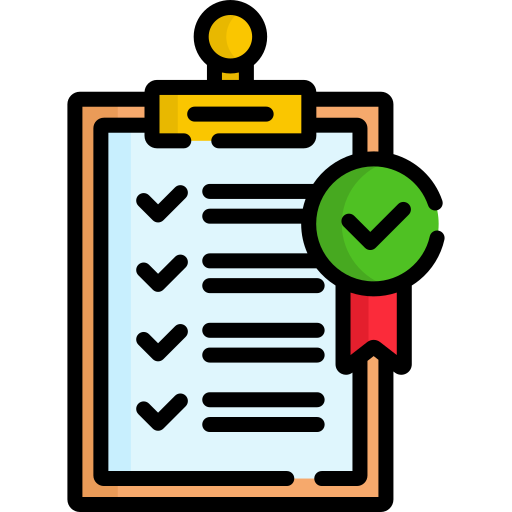}};
  \draw[->, thick] ([xshift=-0.3cm,yshift=0.2cm]tree-left.west) -- ([yshift=-0.05cm]profile.east) node[midway,above,yshift=0.5mm] {\textcolor{orange}{End}};

  \end{tikzpicture}
  %\vspace{-2mm}
  \caption{Overview of the profiling workflow}
  %\vspace{-5mm}
  \label{fig:workflow}
\end{figure}

\hyperref[sec:traffic-capture]{Step 1} consists in performing the interaction with the device under investigation and recording the network traffic. To this end, we connect the Smart Home device (or, for non-IP devices, their gateway or hub) and the device running the triggering source to a wireless (resp. wired) LAN and capture the traffic on the Access Point (AP) (resp. switch). We call an interaction and the resulting device state an \emph{event}. 
%% FDK: commented out because already explained in detail in the next subsection
%We believe, guided by previous work on device profiling \cite{ping-pong}, that the traffic of the Smart Home device and of the controlling device triggering the event are of equal importance, therefore we collect all their traffic. Step 1 is repeated multiple times.

In \hyperref[sec:sig-ex]{step 2}, the traffic traces obtained are filtered and analyzed, and an \emph{event signature} is extracted. The latter consists of a set of traffic flow descriptors, called \emph{Flow IDs} in the following, that describe the relevant features of the bidirectional flows observed in the traces, and therefore associated with the event. A Flow ID contains network and transport layer information, such as host addresses and port numbers, as well as application layer information (when available). The extracted Flow IDs are stored in a tree-shaped data structure that allows tracking which Flow IDs have already been observed.

In \hyperref[sec:event-tree]{step 3}, we select a Flow ID from the tree, configure a firewall running on the switch or AP to block packets matching that ID, and repeat steps 1 and 2. The idea is that by deliberately blocking some of the traffic associated with the interaction and repeating the steps 1 to 2, the device and/or app will try alternative communication patterns.

% until the event fails, i.e., the desired change of the device is not achieved because required communication has been blocked, or 

The steps above are repeated until no new patterns are discovered.
The final result is the profile of the device for the investigated event.
It takes the form of a tree,
where each node is a Flow ID linked to the event,
and the path from the root towards a given node gives
the set of parent Flow IDs which were blocked to let the current Flow ID appear.
%i.e. all the event signatures linked to that event together with the information which flows were blocked to let them appear
We define the Flow IDs initially occurring without any traffic blocking required as \emph{first-level} Flow IDs,
whereas the ones appearing as a result of traffic blocking are \emph{hidden} Flow IDs.
The latter are the communication patterns which can be discovered
only through our multi-level approach,
and therefore missed by state-of-the-art techniques.

\subsection{Step 1: Traffic capturing}
\label{sec:traffic-capture}

Similarly to \emph{PingPong} \cite{ping-pong} and \emph{IoTAthena} \cite{wan_iotathena_2022}, we start capturing the network traffic, perform the interaction, then stop the traffic capture after a predefined duration $d$. We repeat each interaction and the traffic capturing $m$ times.
Our framework automates interactions triggered by a companion app by leveraging the Android Debug Bridge (ADB) tool \cite{adb} to generate touch events on the smartphone running the companion app.
The only manual configuration required by the framework is to plug in the device under test and to connect it to the LAN,
as well as setting up the framework with the smartphone's screen coordinates to issue the touch events.
We also consider power-cycling the Smart home device as an interaction (called \emph{boot interaction} in the following). To automate the boot interaction, we plug the device in a smart power outlet controlled by our framework.

We collect all incoming and outgoing traffic of the Smart Home device and of the smartphone, since we consider the activities of both of them potentially relevant for the characterization of the event. Between captures, we introduce random waiting intervals to avoid the systematic capturing of periodic background traffic unrelated to the interaction.
%(lines~\ref{capturestart}--\ref{captureend} in Algorithm~\ref{algo:exp}).
Control plane packets are filtered out, in particular all ARP, ICMP, DHCP packets, TCP Handshake packets, and TLS Handshake packets, except \texttt{Client Hello} packets containing the Server Name Indication (SNI) extension.

As our approach is based on blocking traffic, it can happen that the interaction does not cause the desired state change in the Smart Home device. For the event signature extraction described in step 2, it is important to filter out such unsuccessful event executions. We do this by verifying the device state, e.g. whether the lamp has turned on, after each capture.
To obtain the device state,
%we leverage the device's API.
we leverage third-party libraries designed to control the device,
e.g. \texttt{python-kasa} \cite{python-kasa} for the TP-Link HS110 plug,
while ensuring this does not interfere with the traffic related to the studied interaction.
We believe this method is more precise and reliable than the approach followed by Mandalari \textit{et al.} \cite{blocking-without-breaking} which compares screenshots of the companion app to check whether the displayed status of the Smart Home device has changed.
If we did not find such a library, or we failed to include it in our framework,
we fall back to screenshots, too, but we compute the Structural Similarity Measure (SSIM) \cite{ssim} and consider two screenshots identical if the SSIM is over an empirically defined threshold, instead of directly comparing image pixels like Mandalari \textit{et al.} The result are $m^+ \leq m$ traffic captures of successful event executions.

% If the device's state was correctly updated,
% we consider the related network traffic to build the event signature.
% Otherwise, we dismiss the captured traffic,
% and proceed to the next event iteration.

\subsection{Step 2: Event signature extraction}
\label{sec:sig-ex}

For each of the $m^+$ packet traces obtained in step 1, the recorded packets are first aggregated to Flow IDs, and then the event signature is extracted. This happens in multiple steps that we now describe.
%(line~\ref{algoextract} in Algorithm~\ref{algo:exp}).

\subsubsection{Replacing IP addresses by domain names}

During an event, the Smart Home device or the companion app may communicate with cloud services provided by the device vendor or other external hosts.
Where possible, we replace non-local source and destination IP addresses by domain names. This makes the signature extraction more robust against changes in the IP addresses caused by cloud migrations or load balancing. To do so, we keep a table matching encountered domain names with their corresponding IP address(es). We first populate the table with entries from the LAN gateway's DNS cache, and then update it whenever a packet bearing domain data is observed, i.e. a DNS query/response or a TLS \texttt{Client Hello} message with the SNI extension.\footnote{We also tried reverse DNS lookups, but since most of the servers contacted during our experiments were hosted at big cloud providers such as Amazon, no meaningful data was obtained, and we abandoned this approach.}
%Therefore, we did not further consider reverse lookups.}

%\begin{itemize}
    %\item For DNS responses, we extract the domain name from the question resource record,
    %along with the related IP address(es) from the answer and additional resource records,
    %and save them in the table.
    %\item For TLS Client Hello packets with the SNI extension,
    %we extract the domain name from said extension,
    %and the IP address for the packet's IP layer.
%\end{itemize}

\subsubsection{Aggregating packets to Flow IDs}

Packets that share all attributes in the set of properties below are grouped to a bidirectional flow, and the attributes form the \emph{Flow ID} of that flow. %We use the following attributes:
\begin{itemize}
\item The hostname (domain name or IP address) of the \emph{source} and \emph{destination}. The Flow ID's \emph{source} is the source of the first packet in the bidirectional flow.
\item The \emph{protocol} (e.g. TCP or UDP).
\item The source and destination \emph{ports}. We only consider a port if it belongs to a well known service (e.g. port 80) or if it appears, for the given combination of the other attribute values, in all $m^+$ packet traces. This means, for example, that the packets of all TCP connections from a client $A$ to the port 80 of server $B$ are aggregated to a single flow and the random client ports used by $A$ are ignored and not further considered. The same is also done for vendor-specific ports, thanks to the above rule.
\item \emph{Application}-specific data for known application layer protocols, amongst others: query name and query type for DNS; method and URI for HTTP; message type, code and URI for CoAP. Consequently, for example, a DNS response is grouped with its query (based on the query type and name).
\end{itemize}

%Grouping along the application-layer data is not as strict as the other attributes,
%as, intuitively, varying application-layer data can be related to the same network flow.
%Therefore, we implemented custom, protocol-specific equality for application-layer data.
%For instance, as a pair of a DNS request and response must be grouped to the same Flow ID,
%we match them on the query type (e.g. A or AAAA) and queried name,
%but not on the QR flag (which indicates if the message is a query or a response).
%Other supported protocols include HTTP and CoAP, for which we respectively match on the method and URI fields,
%and the type (Confirmable or Non-confirmable), code (analogous to HTTP's method), and URI path.

\subsubsection{Building the event signature}

The aggregation of packets to Flow IDs produces a set $f_i$ of Flow IDs, with $1\leq i \leq m^+$, for each of the $m^+$ packet traces.
We define the \emph{event signature} $s$ for the investigated interaction as the set of Flow IDs that appear in all packet traces:
\[
    s := \bigcap_{i=1}^{m^+} f_i
\]
By only keeping the intersection, we filter out Flow IDs belonging to network communications that are not deterministically associated with the event, such as periodic or sporadic messages, as discussed in Section~\ref{sec:problem} and subsection~\ref{sec:traffic-capture}.

\subsection{Step 3: Blocking flows with the event signature tree}
\label{sec:event-tree}

The event signature tree is a \emph{connected}, \emph{acyclic}, \emph{rooted tree} where the nodes of the tree are Flow IDs, except for the root node. The tree is used to keep track of the Flow IDs already seen and blocked.

%The intuition behind the tree is the following: After interacting with the device, we obtain an event signature consisting of Flow IDs. We then pick a Flow ID from the signature and configure the firewall to block packets matching it and repeat the interaction. To keep track which Flow IDs have been already blocked and which still have to be investigated, we use the tree.

\subsubsection{Tree creation}

At the beginning, when the profiling starts for a specific interaction of the device, the tree only consists of the root node. The firewall running on the switch or AP to which the Smart Home device and the companion app are connected does not block any traffic.
%(lines~\ref{algoempty1}--\ref{algoempty2} in Algorithm~\ref{algo:exp}).

After obtaining the event signature $s$ from step 2, we add each Flow ID in the signature as a child node to the root node.
%(lines~\ref{addloop1}--\ref{addloop2} in Algorithm~\ref{algo:exp}).
For the next iteration of steps 1 through 3, we select a node, i.e. a Flow ID, that has not been visited yet from the tree and instruct the firewall to block all packets matching that node and all its ancestors in the tree. The node will be marked as visited and the newly found Flow IDs will be added as children to it, and the procedure is repeated.
%(lines~\ref{algoselect1}--\ref{algoselect2}).

The algorithm ends if no unvisited nodes are left. Theoretically, the algorithm could continue to run indefinitely if new Flow IDs are discovered in each iteration. However, in our experiments, the algorithm always terminated after a finite number of iterations because the more Flow IDs are blocked, the more constrained the communication of the device becomes and the fewer successful event traces are captured, until no more new Flow IDs are added to the tree.

\subsubsection{Tree traversal and pruning}
\label{sec:tree-pruning}

To select the next node,
%(line~\ref{algoselect})
we use a Breadth-First Search (BFS) \cite{algorithms-book} traversal,
i.e., we process all nodes at a given depth before proceeding to the next level.
%We chose BFS over DFS (\textbf{D}epth-\textbf{F}irst \textbf{S}earch) because it more closely fits to our experimental strategy.
Our objective is to trigger all possible network flows corresponding to a Smart Home device event, i.e., express all possible nodes in the event signature tree.
To manage resource usage and execution time, two strategies are used to limit the size of the tree.

Firstly, we already remove event signatures for which less than half of the trace captures were successful, i.e., $m^+ < m/2$, in step 2.
%(lines~\ref{algotest1}--\ref{algotest2} in Algorithm~\ref{algo:exp}).
In this way, we filter out event signatures for which there is not enough data to reliably determine the Flow IDs that are with a high probability linked to the event. In practice, we observe a polarization in terms of successful event execution:
either all event executions succeed ($m^+=m$) or none ($m^+=0$).
The latter occurs when the firewall rules prevent the event's success.
Setting the threshold at $m/2$ is therefore a conservative choice.

\begin{figure}
  \centering
  \begin{tikzpicture}

    % Root
    \node (root) {root};
  
    %% Depth 1
    % Nodes
    \node [anchor=west, right=0.3cm of root, yshift=1cm] (1-plug-phone-tcp) {\circled{A} plug$\leftrightarrow$phone:9999 {[TCP]}};
    \node [anchor=west, right=0.3cm of root, yshift=-0.3cm] (1-plug-server-https) {\circled{B} plug$\leftrightarrow$\texttt{use1-api.tplinkra.com}:443 {[HTTPS]}};
    % Edges
    \draw (root.east) -- (1-plug-phone-tcp.west);
    \draw (root.east) -- (1-plug-server-https.west);
  
    %% Children of 1-plug-phone-tcp
    \node [below=0.4cm of 1-plug-phone-tcp.west, anchor=west, xshift=0.5cm] (A1) (2-plug-server-https) {\circled{C} plug$\leftrightarrow$\texttt{use1-api.tplinkra.com}:443 {[HTTPS]} \xmark};
    \node [below=0.4cm of 2-plug-server-https.west, anchor=west] (2-plug-phone-udp_A) {\circled{D} plug$\leftrightarrow$phone:9999 {[UDP]}};
    \draw ([xshift=0.3cm,yshift=-2.5mm]1-plug-phone-tcp.west) |- (2-plug-server-https.west);
    \draw ([xshift=0.3cm,yshift=-4mm]1-plug-phone-tcp.west) |- (2-plug-phone-udp_A.west);

    %% Children of 1-plug-server-https
    \node [below=0.4cm of 1-plug-server-https.west, anchor=west, xshift=0.5cm] (2-plug-phone-tcp) {\circled{E} plug$\leftrightarrow$phone:9999 {[TCP]} \xmark};
    \node [below=0.4cm of 2-plug-phone-tcp.west, anchor=west] (2-plug-phone-udp_B) {\circled{F} plug$\leftrightarrow$phone:9999 {[UDP]} \xmark};
    \draw ([xshift=0.3cm,yshift=-2.5mm]1-plug-server-https.west) |- (2-plug-phone-tcp.west);
    \draw ([xshift=0.3cm,yshift=-4mm]1-plug-server-https.west) |- (2-plug-phone-udp_B.west);
  
  \end{tikzpicture}
  \caption{\emph{Node pruning}. The nodes \circled{C}, \circled{E}, and \circled{F} are not further explored (\xmark) in the BFS traversal because nodes with the same Flow ID have been already seen (nodes \circled{B}, \circled{A}, and \circled{D}, respectively).}
  \label{fig:tree-pruning}
\end{figure}

Secondly, we prune branches of the tree
which will likely not provide new information,
i.e. no new Flow IDs,
by applying an \textit{ad hoc} pruning heuristic.
To determine which heuristic we can apply without losing any information,
we ran preliminary experiments with one of our testbed's devices,
the TP-Link HS110 smart plug \cite{hs110}.
We observed that, for any two nodes with the same Flow ID,
their children were always identical.
We conclude it is likely that no new information will be found
by processing a node identical to one which has already been processed,
regardless of its position in the tree.
Therefore, we adopt the \emph{node pruning} heuristic,
illustrated in Fig.~\ref{fig:tree-pruning}:
a node is pruned if an equivalent node, i.e., a node with the same Flow ID, has already been processed.
Additional insight concerning this preliminary experiment is given in appendix \ref{app:no-pruning}.

\subsubsection{How to block packets matching a Flow ID}\label{sec:firewall}

As explained, a Flow ID comprises traffic features from multiple networking layers, such as hostnames, port numbers, and application layer information. 
The presence of application layer information in the Flow ID requires a firewall that is able to match and reject traffic based on information found at that layer. A simple firewall, such as the default Linux kernel firewall NFTables \cite{nftables} (successor of the well-known IPTables) is not sufficient.
Instead, we leverage the open-source Smart Home firewall we developed as previous work \cite{smart-home-firewall}.
It is based on NFTables, while enhancing its capabilities to match additional application-layer protocols.
However, this firewall can only work with \emph{allow} rules, while our profiling algorithm is based on blocking rules.
Therefore, we modified the firewall, turning it into a \emph{deny-list} firewall. 
In our prototype implementation of the profiling algorithm, we translate the list of Flow IDs into firewall blocking rules and let the firewall enforce them for the next batch of experiments.
The code of the modified firewall is available at \url{https://github.com/smart-home-network-security/firewall-blocklist}.

\section{Experimental results}
\label{sec:results}

In this section, we present our experimental setup,
and the results obtained from applying our framework to it.
%We assess the device's network communication robustness, at boot and for each type of interaction with the device.
% For the results, we first show an example of a complete event signature tree
% generated by our framework,
% then present metrics, computed over all the obtained trees,
% with the goal of assessing each event's robustness.

%% Legacy paragraph, from when the paper's focus was not really over robustness
% Every event showcasing new communication flows at a tree depth higher than 1
% backs up our initial claim that devices provide hidden communication patterns,
% which must be taken into account for an exhaustive device profiling.
% Additionally, it shows that the devices reveal some robustness to network instability,
% as events might still succeed if its first-level traffic is unsuccessful.

\subsection{Experiment setup}

We apply our methodology to a controlled testbed network comprised of real-world devices,
mimicking a typical Smart Home network, depicted in Fig.~\ref{fig:testbed}.
We examine an array of commercial, off-the-shelf Smart Home devices,
including four power plugs, three cameras, and three light bulbs.
The profiled interactions, per device category, are the following:
\begin{itemize}
  \item For all devices: booting the device;
  \item Power plugs: toggling them on/off;
  \item Cameras: streaming their video feed;
  \item Light bulbs: toggling them on/off, changing their brightness, changing their color.
\end{itemize}

\begin{figure}
    \centering
    
    \begin{tikzpicture}[node distance=2cm,on grid,auto]
  
        %%% NODES %%%
      
        % Routers
        \node (ap){\includegraphics[scale=0.06]{figures/testbed/wireless-ap.png}};
        \node (router)[right=1.3cm of ap,yshift=1cm]{\includegraphics[scale=0.07]{figures/testbed/router.png}};
  
        %% IP devices
        \node (phone) [right=1.3cm of ap,yshift=0cm] {\includegraphics[scale=0.05]{figures/testbed/smartphone.png}};
        % Plugs
        \node (wifi-plug-1)[left=2.5cm of ap, yshift=8mm]{\includegraphics[scale=0.035]{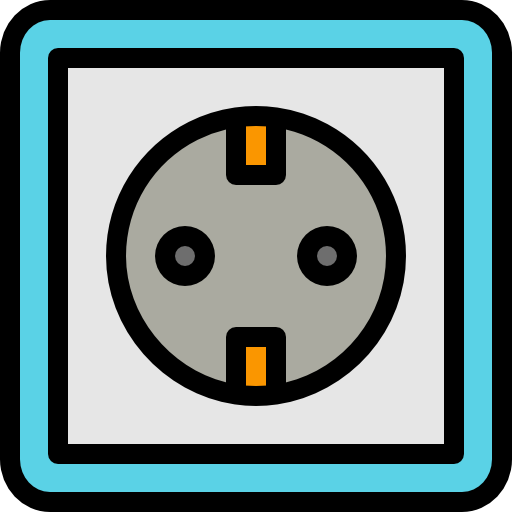}};
        \node (wifi-plug-2)[left=0.6cm of wifi-plug-1]{\includegraphics[scale=0.035]{figures/testbed/power-socket.png}};
        \node (wifi-plug-3)[left=0.6cm of wifi-plug-2]{\includegraphics[scale=0.035]{figures/testbed/power-socket.png}};
        % Cameras
        \node (camera-1)[left=2.5cm of ap]{\includegraphics[scale=0.05]{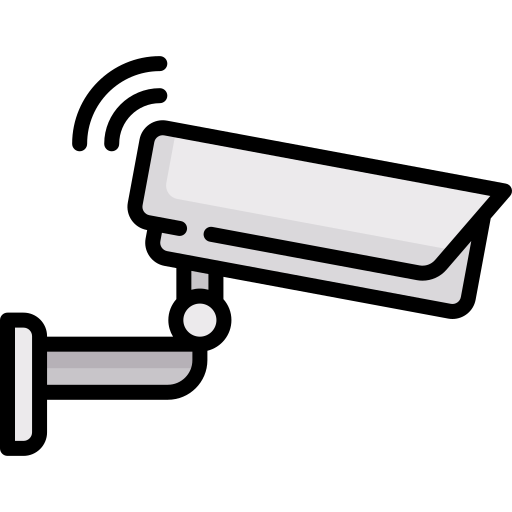}};
        \node (camera-2)[left=0.8cm of camera-1]{\includegraphics[scale=0.05]{figures/testbed/camera.png}};
        \node (camera-3)[left=0.8cm of camera-2]{\includegraphics[scale=0.05]{figures/testbed/camera.png}};
        % Lights
        \node (wifi-light-1)[left=2.5cm of ap, yshift=-8mm]{\includegraphics[scale=0.05]{figures/testbed/light-bulb.png}};
        \node (wifi-light-2)[left=0.7cm of wifi-light-1]{\includegraphics[scale=0.05]{figures/testbed/light-bulb.png}};
  
        % Hubs / Gateways
        \node (hue-bridge)[below=1cm of ap]{\includegraphics[scale=0.05]{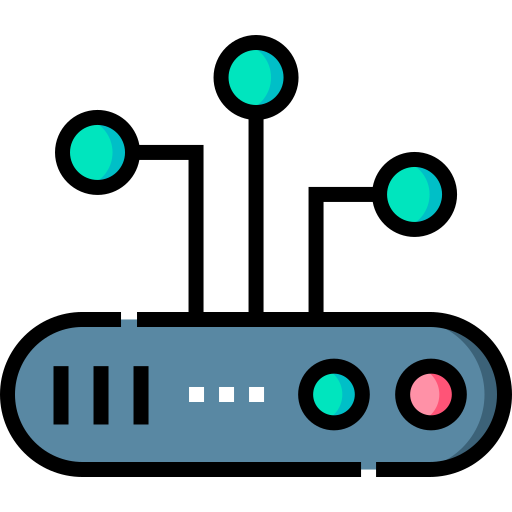}};
        \node (st-hub)[above=1cm of ap]{\includegraphics[scale=0.05]{figures/testbed/gateway.png}};
  
        % Zigbee devices
        \node (hue-light)[left=1.2cm of hue-bridge]{\includegraphics[scale=0.05]{figures/testbed/light-bulb.png}};
        \node (st-plug)[left=1.2cm of st-hub]{\includegraphics[scale=0.035]{figures/testbed/power-socket.png}};
  
        % Internet
        \node (internet)[right=1.7cm of router]{\includegraphics[scale=0.08]{figures/testbed/cloud.png}};

        %%% EDGES %%%
  
        % WAN
        \draw [wan]    (ap)          -- (router);
        \draw [wan]    (router)      -- (internet);
  
        % Wired LAN
        \draw [lan]    ([yshift=-0.1cm]hue-bridge.north)  -- ([yshift=0.1cm]ap.south);
        \draw [lan]    ([yshift=0.1cm]st-hub.south) -- ([yshift=-0.1cm]ap.north);
  
        % WLAN
        \draw [wlan] (phone)       -- (ap);
        \draw [wlan] (wifi-plug-1)  -- (ap);
        \draw [wlan] (camera-1)    -- (ap);
        \draw [wlan] (wifi-light-1)  -- (ap);
  
        % Zigbee
        \draw [zigbee]   (hue-light)  -- (hue-bridge);
        \draw [zigbee]   (st-plug)     -- (st-hub);

        %%% LEGEND %%%
  
        % Caption
        \node (legend)  [below=7.5mm of internet,xshift=-6mm] {\textbf{Legend}};
        % WAN
        \node (wan-a)   [below=0.1cm of legend.south west] {};
        \node (wan-b)   [right=0.6cm of wan-a] {};
        \draw [wan]     (wan-a) -- (wan-b);
        \node (wan-key) [right=0.4cm of wan-a.east] {WAN};
        % Wired LAN
        \node (lan-a)   [below=0.3cm of wan-a] {};
        \node (lan-b)   [right=0.6cm of lan-a] {};
        \draw [lan]     (lan-a) -- (lan-b);
        \node (lan-key) [right=0.4cm of lan-a.east] {Wired LAN};
        % WLAN
        \node (wlan-a) [below=0.3cm of lan-a] {};
        \node (wlan-b) [right=0.6cm of wlan-a] {};
        \draw [wlan]   (wlan-a) -- (wlan-b);
        \node (wlan-key) [right=0.4cm of wlan-a.east] {WLAN};
        % Zigbee
        \node (zigbee-a) [below=0.3cm of wlan-a] {};
        \node (zigbee-b) [right=0.6cm of zigbee-a] {};
        \draw [zigbee]   (zigbee-a) -- (zigbee-b);
        \node (zigbee-key) [right=0.4cm of zigbee-a.east] {Zigbee};
        % Box
        \coordinate [above=0.25cm of legend,xshift=-0.65cm] (rect-top-left);
        \coordinate [below=0.25cm of zigbee-key,xshift=1.15cm] (rect-bottom-right);
        \draw [ultra thin] (rect-top-left) rectangle (rect-bottom-right);
  
    \end{tikzpicture}
    %\vspace{-2mm}
    \caption{Experimental Smart Home network.}
    %\vspace{-5mm}
    \label{fig:testbed}
  \end{figure}

%% Old paragraph
% the default usage scenario involves controlling the device under test
% through its official companion app;
% nonetheless, it is also possible to leverage a \emph{Smart Home automation platform},
% which can control the device on behalf of the official app.
% To account for both scenarios,
% we derive up to three usage scenarios per device:

In a real Smart Home network,
a user could choose among various apps to control their devices,
including the device's official companion app (from its manufacturer or vendor),
a third-party app,
or a Smart Home automation platform which provides
unified control over devices from different manufacturers.
Such different home automation methods can trigger diverging, but valid, communication patterns.
To cover such cases,
for all interactions other than booting,
we derive up to three usage scenarios:
%to account for various ways of controlling the device:
\begin{itemize}
  \item For all devices, we issue their interactions using their official companion app;
  \item For devices that support it, we also control them using the app of the popular home automation platform \emph{SmartThings}~\cite{smartthings};
  \item For the TP-Link plug and the Hue light bulb, we also experiment with other apps; respectively, \emph{TP-Link Tapo} \cite{app-tapo} (another app from the same manufacturer, targeted toward \emph{Tapo}-branded devices) and \emph{Hue Essentials} \cite{app-hue-essentials} (third-party app for Hue devices).
\end{itemize}

The smartphone running the device-specific apps is a Crosscall CORE-X4 \cite{phone}
with the Android 10 OS.

\begin{table}
  \centering
  \begin{tabular}{|c|c|c|c|}
    \hline
    \textbf{Device} & \textbf{App} & \textbf{Interaction} & \textbf{Event ID} \\
    \hline
    \multicolumn{4}{|c|}{\textbf{Power plugs}} \\
    \hline
    \multirow{4}{*}{\makecell{TP-Link HS110\\\cite{hs110}}} & / & boot & 1 \\
    & Kasa Smart & toggle & 11 \\
    & TP-Link Tapo & toggle & 12 \\
    & SmartThings & toggle & 13 \\
    \hline
    \multirow{2}{*}{\makecell{SmartThings Outlet $\dag$\\\cite{st-outlet}}} & / & boot & 2 \\
    & SmartThings & toggle & 14 \\
    \hline
    \multirow{3}{*}{\makecell{Tapo P110\\\cite{tapo-p110}}} & / & boot & 3 \\
    & TP-Link Tapo & toggle & 15 \\
    & SmartThings & toggle & 16 \\
    \hline
    \multirow{2}{*}{\makecell{Woox R5024 (Tuya)\\\cite{woox-plug}}} & / & boot & 4 \\
    & Tuya Smart & toggle & 17 \\
    \hline
    \multicolumn{4}{|c|}{\textbf{Cameras}} \\
    \hline
    \multirow{2}{*}{\makecell{Xiaomi MJSXJ02CM\\\cite{xiaomi-cam}}} & / & boot & 5 \\
    & Mi Home & stream & 18 \\
    \hline
    \multirow{3}{*}{\makecell{Tapo C200\\\cite{tapo-c200}}} & / & boot & 6 \\
    & TP-Link Tapo & stream & 19 \\
    & SmartThings & stream & 20 \\
    \hline
    \multirow{2}{*}{\makecell{D-Link DCS-8000LH\\\cite{dlink-cam}}} & / & boot & 7 \\
    & mydlink & stream & 21 \\
    \hline
    \multicolumn{4}{|c|}{\textbf{Light bulbs}} \\
    \hline
    \multirow{8}{*}{\makecell{Philips Hue White and\\Color lamp A60 $\dag$\\\cite{hue-lamp}}} & / & boot & 8 \\
    & Philips Hue & toggle & 22 \\
    & Hue Essentials & toggle & 23 \\
    & Hue Essentials & brightness & 24 \\
    & Hue Essentials & color & 25 \\
    & SmartThings & toggle & 26 \\
    & SmartThings & brightness & 27 \\
    & SmartThings & color & 28 \\
    \hline
    \multirow{3}{*}{\makecell{Alecto Smart-Bulb10\\(Tuya) \cite{alecto-smart-bulb}}} & / & boot & 9 \\
    & Tuya Smart & toggle & 29 \\
    & Tuya Smart & color & 30 \\
    \hline
    \multirow{7}{*}{\makecell{Tapo L530E\\\cite{tapo-l530e}}} & / & boot & 10 \\
    & TP-Link Tapo & toggle & 31 \\
    & TP-Link Tapo & brightness & 32 \\
    & TP-Link Tapo & color & 33 \\
    & SmartThings & toggle & 34 \\
    & SmartThings & brightness & 35 \\
    & SmartThings & color & 36 \\
    \hline
  \end{tabular}
  \caption{
    Devices in our testbed, with corresponding apps and interactions.
    $\dag$ indicates Zigbee devices, for which the hub traffic was considered.
  }
  \label{tab:devices_id}
\end{table}

The complete device references are listed in Table~\ref{tab:devices_id},
along with the corresponding smartphone apps and interactions instrumented.
The numerical identifier used as x-axis for some subsequent plots
pertains to the \emph{Event ID} listed in the table,
which is a unique number assigned by us to each of the tested combinations of device, app, and interaction.
As explained, we only consider IP traffic,
whether over Ethernet or Wi-Fi.
Therefore, for the two devices communicating using Zigbee,
we profile the traffic issued by their hub,
i.e. the Hue bridge for the Hue lamp,
and the SmartThings hub for the SmartThings Outlet,
respectively.
For the \emph{boot} interaction experiments over those devices,
we synchronously power-cycle both the device itself and its hub.

Our practical experimental parameters, inspired by related works \emph{PingPong} \cite{ping-pong} and \emph{IoTAthena} \cite{wan_iotathena_2022}, are the following:
\begin{itemize}
  \item Number of event iterations: $m = 20$;
  \item Traffic capture timeout: $d = 20$ seconds (90 seconds for interaction \emph{boot});
  % \item Number of event iterations (defined as $m$ in Section~\ref{sec:traffic-capture}): $m = 20$;
  % \item Traffic capture timeout (defined as $d$ in Section~\ref{sec:traffic-capture}): $d = 20$ seconds (90 for event \emph{boot});
  \item Duration between two event iterations: randomly chosen between 40 (120 for interaction \emph{boot}) and 150 seconds.
\end{itemize}

\subsection{Example signature tree}

As an example of our framework's results,
we present the event signature tree
generated for one device of our corpus,
namely the TP-Link HS110 \cite{hs110}.
This device being a smart power plug,
it provides one interaction, i.e. toggling it on and off.
For this experiment, we controlled the device using its official Android companion app,
i.e. Kasa Smart \cite{app-kasa}. The resulting signature tree
is displayed in Fig.~\ref{fig:device-tree}.
To avoid unnecessarily cluttering the figure,
we only show the first occurrence of each unique flow.
%% Old version
%In general, every \emph{first-level} flow occurs as a child of every tree node.
%% New version
We empirically observe that each one of the listed \emph{first-level} flows
also occurs as a child of every other node in the tree,
i.e. the communication pattern is relatively stable.
In certain cases, a \emph{hidden} flow will appear at deeper levels,
which embodies a backup communication strategy
when the default one fails.
%\CP{Are there many repeating flows?}
%\FDK{A lot. Basically, each first-level flow appears as a child of every other.}
%\CP{We need to say something about this.}
%\FDK{Reworked the paragraph to address this.}

%\newcommand*\circled[1]{\raisebox{.5pt}{\textcircled{\raisebox{-.9pt} {#1}}}}
\begin{figure}
  \centering
  \begin{tikzpicture}

    % Root
    \node (root) {root};
  
    %% Depth 1
    % Nodes
    \node [anchor=west, right=0.4cm of root, yshift=-0.25cm] (D) {\circled{D} phone$\rightarrow$broadcast(\texttt{255.255.255.255}):9999 {[UDP]}};
    \node [above=0.4cm of D.west,anchor=west] (C) {\circled{C} phone$\rightarrow$mDNS(\texttt{224.0.0.251}):5353 {[UDP / mDNS]}};
    \node [above=0.4cm of C.west,anchor=west] (B) {\circled{B} phone$\leftrightarrow$plug:9999 {[UDP]}};
    \node [above=1.1cm of B.west,anchor=west] (A) {\circled{A} phone$\leftrightarrow$plug:9999 {[TCP]}};
    \node [below=1.1cm of D.west,anchor=west] (E) {\circled{E} phone$\leftrightarrow$\texttt{xx-device-telemetry}...:443 {[HTTPS]}};
    \node [below=0.4cm of E.west,anchor=west] (F) {\circled{F} plug$\leftrightarrow$gateway:53 {[UDP / DNS: A \emph{tplinkapi}]}};
    \node [below=0.4cm of F.west,anchor=west] (G) {\circled{G} plug$\leftrightarrow$\texttt{use1-api.tplinkra.com}:443 {[HTTPS]}};
    % Edges
    \draw (root.east) -- (A.west);
    \draw (root.east) -- (B.west);
    \draw (root.east) -- (C.west);
    \draw (root.east) -- (D.west);
    \draw (root.east) -- (E.west);
    \draw (root.east) -- (F.west);
    \draw (root.east) -- (G.west);
  
    %% Children of A
    \node [below=0.35cm of A.west, anchor=west, xshift=0.5cm] (A1) {plug$\leftrightarrow$\texttt{79.125.56.92}:443 {[HTTPS]}};
    \node [below=0.35cm of A1.west, anchor=west] (A2) {phone$\leftrightarrow$\texttt{n-wap.tplinkcloud.com}:443 {[HTTPS]}};
    \draw ([xshift=0.3cm,yshift=-2.5mm]A.west) |- (A1.west);
    \draw ([xshift=0.3cm,yshift=-0.35cm]A.west) |- (A2.west);
  
    %% Children of D
    \node [below=0.35cm of D.west, anchor=west, xshift=0.5cm] (D1) {plug$\leftrightarrow$\texttt{34.240.186.173}:443 {[HTTPS]}};
    \node [below=0.35cm of D1.west, anchor=west, xshift=0.5cm] (D2) {plug$\leftrightarrow$\texttt{n-devs.tplinkcloud.com}:443 {[HTTPS]}};
    \draw ([xshift=0.3cm,yshift=-2.5mm]D.west) |- (D1.west);
    \draw ([xshift=0.3cm,yshift=-2mm]D1.west) |- (D2.west);
  
    \end{tikzpicture}
  \caption{
    Event signature tree for the TP-Link HS110's \emph{toggle} event.
    $\leftrightarrow$ (resp. $\rightarrow$) denotes bidirectional (resp. unidirectional) flows.
    When indicated, the port is associated with its host.} 
  \label{fig:device-tree}
\end{figure}

By default, toggling the plug triggers seven different flows:
\begin{itemize}
  \item[\circled{A}] between the phone and the plug's TCP port 9999;
  \item[\circled{B}] between the phone and the plug's UDP port 9999;
  \item[\circled{C}] from the phone to the mDNS multicast address (224.0.0.251, UDP port 5353);
  \item[\circled{D}] from the phone to the broadcast address' (255.255.255.255) UDP port 9999;
  \item[\circled{E}] HTTPS between the phone and the server \texturl{xx-device-telemetry-gw.iot.i.tplinknbu.com};
  \item[\circled{F}] DNS query/response from the plug to the LAN gateway, for the domain name \texturl{use1-api.tplinkra.com};
  \item[\circled{G}] HTTPS between the plug and the server \texturl{use1-api.tplinkra.com}.
\end{itemize}
Flows \circled{C} and \circled{D} are unidirectional,
while the remaining ones are bidirectional.

When one of those flows is blocked,
the device might issue different network traffic to perform its event.
Indeed, when the TCP communication \circled{A} on the plug's port 9999 is blocked,
we see two new HTTPS flows appearing:
between the phone and the server \texturl{n-wap.tplinkcloud.com}, and
between the plug and the IPv4 address \texturl{79.125.56.92}.
The former also occurs when other first-level flows are blocked,
including the similar pattern \circled{B} using UDP instead of TCP.
Both flows are part of a backup strategy,
going through the internet,
if the device is not able to communicate locally with the controlling phone.
Such \emph{hidden} backup flows also appear as children of the \emph{first-level} flow \circled{D}.

\subsection{Assessing device event robustness}
\label{sec:robustness}

%% Legacy paragraphs, when the plots were in the body text instead of the appendix

% As a quantitative evaluation of our framework,
% we derived various descriptive metrics over our experimental corpus.
% The first one is the count of unique Flow IDs discovered,
% which represents the variability of our devices' communication patterns.
% This metric was further split into two following groups bearing different semantics:
% the count of unique \emph{first-level} Flow IDs,
% and the count of unique \emph{hidden} Flow IDs.
% The former exhibits the patterns that state-of-the-art techniques could discover,
% as they appear without disturbing the traffic.
% The latter, however, illustrates the patterns which appear at tree depths of two and above,
% and can thus be discovered only through our multi-level approach.
% Fig.~\ref{fig:count_flow_id} depicts such metrics;
% the first 10 events being the \emph{boot} events,
% and the subsequent ones the remaining user-triggered events.

% The count of unique \emph{hidden} Flow IDs
% can be used as an indicator of the robustness of the device.
% Indeed, it embodies the alternative communication strategies
% that a device might use if the default one fails,
% effectively enhancing the device's robustness to network instability.
% As such, we rebrand this number as the \emph{robustness score}.
% Fig~\ref{fig:robustness-score} depicts this count across all device events,
% as well as its mean, which is 1.94.

The experiments in our testbed of devices result in
the extraction of 254 unique Flow IDs,
of which 70 are \emph{hidden} (i.e. 27.56\%).
Their distribution over the 36 tested event configurations is shown in Fig.~\ref{fig:count_flow_id} in the appendix.
\emph{Hidden} Flow IDs embody the alternative communication strategies
that a device might use if the default one fails,
effectively enhancing the device's robustness to network instability (remember that we only keep Flow IDs associated with successful event executions).
We define the robustness score as the count of hidden Flow IDs.
Recall that a \emph{hidden} Flow ID is one which only appears in the tree deeper than the first level, i.e. resulting from blocking specific flows.
We compute this metric for each tested event;
Fig.~\ref{fig:robustness-score} shows the results.

\begin{figure}
  \centering
  \includegraphics{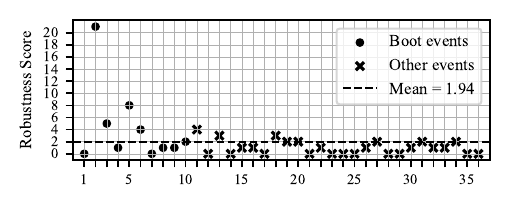}
  \caption{
    \emph{Robustness score}.
    The x-axis shows the \emph{Event ID}.
  }
  \label{fig:robustness-score}
\end{figure}

Out of the 36 instrumented events,
23 (63.9\%) have a robustness score of at least one (mean 1.94).
This value is encouraging, as it means that
most types of interactions dispose of at least one backup strategy
in the case of a failure.

To gain further insight into the distribution of robustness scores,
%and inspired by the categorization done by Hu \textit{et al.} \cite{iot_ipv6} for IPv6 usage behavior,
Fig.~\ref{fig:robustness-grouped} shows the same \emph{robustness score},
grouped along three axes, namely:
\begin{itemize}
  \item Device category;
  \item Controlling app;
  \item Manufacturer.
\end{itemize}

\begin{figure}
  \centering
  \includegraphics{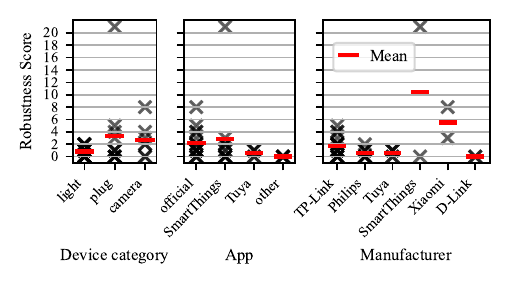}
  \caption{\emph{Robustness score} for all events, grouped per device category, app, and manufacturer.
  Each marker represents the robustness score of one event.
  ST: SmartThings.}
  \label{fig:robustness-grouped}
\end{figure}

We identify an event with a robustness score significantly higher than all others:
the \emph{boot} event for the SmartThings Outlet device.
Being a Zigbee-communicating device,
its \emph{boot} event consists in switching on the plug itself
as well as its related hub.
As hubs are commonly used to provide centralized control over a Smart Home network,
they must provide a way to detect the devices present in the network.
It is expected that, if the default way fails,
the hub will follow backup communication paths to provide this functionality.

We observe that the different device categories in our testbed do not provide the same robustness.
The most robust devices seem to be the power plugs.
A rationale might be that,
as the plugs' operation (switching on or off) is the simplest,
it takes less effort to the manufacturers to provide backup strategies.
However, given the size of the testbed compared to the vast number of devices on the market, this observation should be taken as indicative only. This also applies to the following discussion of the companion apps and manufacturers.

Among the companion apps,
Tuya seems to be less robust than the others.
While most apps provide device control
either via the local network or through the cloud,
Tuya devices are notoriously reluctant to LAN-based control,
preferring cloud-based communication.
This decision reduces their robustness by design,
as, when the cloud endpoints cannot be reached,
the device cannot function properly.
SmartThings boasts a good score,
both as a controlling app and a manufacturer,
helped by a very high robustness score for the SmartThings Outlet's \emph{boot} interaction.
We can also see that the "other" app category provides no backup strategy whatsoever,
suggesting that a user should prefer using trusted and official apps, even if the devices' API is theoretically open.

%\FDK{Added the following paragraph under Cristel's advice.}
We illustrate the effect of the choice of the app with the Philips Hue lamp device and its \emph{toggle} interaction,
which we have triggered by the official app (Philips Hue),
the SmartThings app, and a third-party app (Hue Essentials).
The official app and SmartThings show similar results:
five and six \emph{first-level} Flow IDs respectively,
and one \emph{hidden} flow for each.
However, when using the Hue Essentials app,
only three \emph{first-level} flows are extracted,
and no \emph{hidden} flow.

The Xiaomi camera also stands out with a high mean score.
As cameras are usually tied to more critical functions,
such as monitoring a home against potential intrusions,
Xiaomi followed a respectable rationale here.
TP-Link seems quite robust,
while Philips, even as a trusted device manufacturer,
disappoints in terms of robustness.

% \FDK{This paragraph has been kept from the previous version. Still relevant ?}
% A possible way to increase the number of interactions to investigate, and potentially the number of discovered communication patterns, is to include experimental setups where two or more IoT devices communicate with each other. In most home automation solutions, actions involving multiple devices, such as switching on a lamp by a movement sensor, are centrally managed by a dedicated hub, but some devices on the market support control message exchanges between devices without a hub. We deem as future work to cover this kind of traffic. We argue that our methodology can still be applied, albeit with some modifications: on the one hand, the setup of such device interactions on a home automation platform, and on the other hand a broader traffic filter inside the LAN, to accommodate all devices potentially taking part to the interactions. \CP{rephrase this last sentence.}
%It should however be noted that the signature tree for events involving multiple devices might undergo a state space explosion problem. 

\subsection{A closer look at DNS data}

% A primordial implementation of network communication robustness
% lies within interactions with the DNS protocol.
The Domain Name System (DNS), which allows communicating with hosts
by providing a domain name instead of their IP address, 
is a useful tool to provide application robustness.
% serves network robustness by design:
% a same name can point to different and/or multiple addresses,
% depending on the network's state.
% With this in mind,
Here, we steer our analysis towards interactions with the DNS protocol,
guided by two questions:
\begin{itemize}
  \item If the servers the device tries contacting are unresponsive,
  will it contact servers with other domain names?
  \item If the device's default DNS resolver is unreachable,
  will the device try other ones?
\end{itemize}

We focus on event signatures related to \emph{boot} interactions,
as those proved to provide the most information concerning DNS usage.
Fig.~\ref{fig:count_dns_boot} shows the data related to both questions:
the count of unique domain names contacted,
and that of DNS resolvers queried.
In both cases, the \emph{first-level}
\footnote{
In this context, the qualifier \emph{first-level}
indicates the occurrences at the signature tree's first depth level;
it has no relation with the similarly named, but unrelated, domain names' top-level domain.
}
and the \emph{hidden} count are shown.
% The results for all types of interactions is shown in Fig.~\ref{fig:count_dns_all}
% in the appendix.

\begin{figure}
  \centering
  \stackengine{0.25\linewidth}{% Adjust overlap amount
    \includegraphics{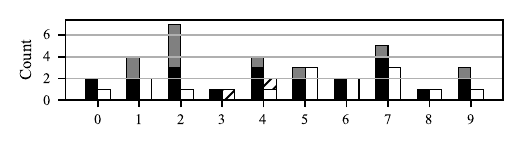}
  }{
    \includegraphics{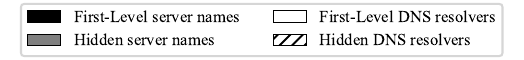}
  }{O}{c}{F}{F}{L}
  \caption{
    Domain names / DNS resolvers contacted for each \emph{boot} event.
    The x-axis shows the \emph{Event ID}.
  }
  \label{fig:count_dns_boot}
\end{figure}

\subsubsection{Server's domain names}

We observe that, out of the 10 \emph{boot} events,
six consider contacting at least one alternative domain name
if the default one fails.

\subsubsection{DNS resolvers}

Out of the 17 DNS resolvers contacted,
only two are \emph{hidden} resolvers.
One of them is simply the LAN gateway (192.168.1.1),
meaning the only real new DNS resolver discovered at a level deeper than one is the Chinese public DNS resolver 114.114.114.114,
contacted by the Xiaomi camera when the LAN gateway is not responsive.
This result shows that, regarding domain name resolving,
Smart Home devices still lack robustness.
Indeed, every device should try contacting a backup resolver if the primary one fails,
as this has been made easy thanks to publicly available resolvers such as Cloudflare's 1.1.1.1 or Google's 8.8.8.8.

\subsection{Analysis of the node pruning heuristic}

We analyze the effect of our tree pruning heuristic
described in Section~\ref{sec:tree-pruning}.
Fig.~\ref{fig:count-pruned} illustrates,
for every studied event,
the count of Flow IDs which have been pruned
while generating the event signature tree,
split per node depth.
As a reminder, this heuristic states that we prune a node, i.e. we do not process the children of a node,
if an equivalent node is already present in the tree;
the depicted count is therefore equal to the number of duplicate Flow IDs in the tree.

\begin{figure}
  \centering
  \includegraphics{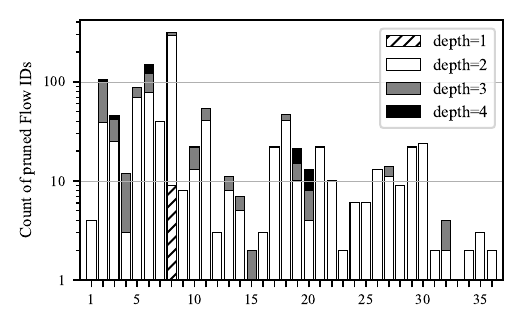}
  \caption{
    Pruned Flow IDs.
    The x-axis shows the \emph{Event ID}.
  }
  \label{fig:count-pruned}
\end{figure}

% Communication redundancy
% Justification for node pruning

We observe that the count is generally important.
In terms of efficiency, this means that our heuristics manages to
significantly reduce the time and resources taken by the event signature tree generation.

We also notice that most of the duplicate Flow IDs occur at depth 2.
Those represent all the flows that occur after blocking a first-level flow.
Their large number shows that most devices exhibit a high redundancy in communication patterns.
Indeed, if the device must change its network behavior to circumvent a traffic restriction,
it will usually only change the affected protocol,
e.g. by contacting another domain name or switching from TCP to UDP; 
it will not completely modify the communication pattern.

\subsection{Comparison with related work}

We compare the profiles generated by our framework
with those generated by methods proposed in related work,
namely \emph{PingPong} \cite{ping-pong},
\emph{BehavIoT} \cite{behaviot},
and \emph{MUDgee} \cite{mudgee},
for the devices present in both their testbeds and ours.
\emph{PingPong} and \emph{BehavIoT} profile specific interactions of a device, 
whereas \emph{MUDgee} profiles the device as a whole.
Table~\ref{tab:devices_coverage} summarizes the devices and interactions covered by each work,
as well as the smartphone apps which produced the subset of our data used for our side in the comparison.

\begin{table}
  \centering
  \begin{tabular}{c|c|c|c|c}
   & & \textbf{PingPong} & \textbf{BehavIoT} & \textbf{MUDgee} \\
	Devices & Apps & \cite{mudgee} & \cite{ping-pong} & \cite{behaviot} \\
  \hline
  TL plug & Kasa Smart & toggle & toggle & device \\
  Hue lamp & Philips Hue & toggle & \xmark & device \\
  ST plug & SmartThings & toggle & boot & device \\
  DL cam & mydlink & \xmark & stream & \xmark \\
  \end{tabular}
  \caption{Related works' device and interaction coverage. Device legend: TL plug = TP-Link HS110; Hue lamp = Philips Hue lamp; ST plug = SmartThings Outlet; DL cam = D-Link camera.}
  \label{tab:devices_coverage}
\end{table}

\begin{figure}
  \centering
  \includegraphics{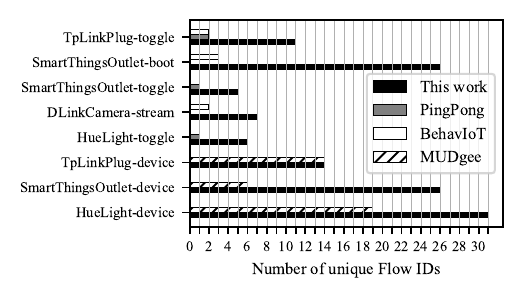}
  \caption{Comparison of discovered unique Flow IDs between three related works and ours. Note that MUDgee profiles devices as a whole and not for individual interactions.}
  \label{fig:comparison_all}
\end{figure}

% \CP{This shows one side of the picture. It is interesting to also mention the number of devices they have that we do not support. Maybe in appendix.}
% \FDK{Not sure if it is really relevant. The appendices' length is limited.}
%\CP{Do you reproduce the other works or cite the results that are expressed in the paper ? If you reproduce and share the code, this is added value for your papers.}
%\FDK{For PingPong and MUDgee, I simply took their reported results. For BehavIoT, I reproduced their experiments. I explained that explicitly.}
Fig.~\ref{fig:comparison_all} compares the number of unique Flow IDs discovered by each work and by us.
Data from \emph{PingPong} and \emph{MUDgee} were taken from their reported results,
whereas \emph{BehavIoT}'s results were reproduced on our side.
The three related works, by design,
do not explore the event signature tree deeper than the first level.
Regarding the specific interactions,
our approach is, as expected, able to extract more Flow IDs
than \emph{PingPong} and \emph{BehavIoT},
thanks to a deeper tree inspection,
and also a more thorough communication pattern extraction, even at the first level.
To compare our results with the device-level profiles created by \emph{MUDgee}, we count the unique Flow IDs over all interactions of a device. For two of the instrumented devices, namely the SmartThings Outlet and the Hue light,
our approach is again able to extract a significantly higher number of Flow IDs than \emph{MUDgee}.
However, the number of Flow IDs is identical for the TP-Link HS110. A manual investigation shows that the extracted Flow IDs are different:
10 out of 14 \emph{MUDgee} flows are NTP traffic towards various servers,
whereas our solution did not extract any NTP patterns.
This traffic might be part of periodic patterns,
which are not covered by our approach;
as explained in Section~\ref{sec:traffic-capture}, our methodology excludes traffic that does not appear in every iteration of a tested interaction from the profile building process.
On the other hand, we extract 10 non-NTP traffic patterns
which were not detected by \emph{MUDgee}.

Additionally, we compare the number of domains names discovered by our approach with those reported by Mandalari \emph{et al.} \cite{blocking-without-breaking}. The latter focuses on extracting the hosts from Smart Home communication patterns.
Table~\ref{tab:iotrimmer} shows, in the columns ``Total ours'' and ``Total theirs'', the total number of unique domain names discovered by our approach and by theirs for the devices and interactions covered in both works.
We observe that our multi-level approach is able to discover more domain names. However, it should be noted that changes in the firmware of the devices since the publication of the cited article \cite{blocking-without-breaking} could influence the results.

\begin{table*}
  \centering
  \begin{tabular}{|c|c|c|c|c|c|}
		\hline
    \textbf{Device \& interact.} & Total ours & Total theirs & Both & Ours only & Theirs only \\
    \hline

    \makecell{
      TP-Link HS110\\
      boot \& toggle
    } &
    5 & 4 &
    \makecell{
      use1-api.tplinkra.com\\
      euw1-api.tplinkra.com
    } & \makecell{
      euw1-device-telemetry-gw.iot.i.tplinknbu\\
      n-wap.tplinkcloud.com\\
      n-devs.tplinkcloud.com\\
    } & n-deventry.tplinkcloud.com \\
    \hline

    \makecell{
      SmartThings Outlet\\
      boot
    } & 4 & 3 & / &
    \makecell{
      hub.smartthings.com\\
      usher.connect.smartthings.com\\
      connectivity.smartthings.com\\
      dc2-eu01-euwest1.connect.smartthings.com
    } & fw-update2.smartthings.com \\
    \hline

    \makecell{
      Philips Hue lamp\\
      boot
    } & 5 & 4 & / &
    \makecell{
      mqtt-eu-01.iot.meethue.com\\
      time.meethue.com\\
      ntp2.aliyun.com\\
      ntp4.aliyun.com\\
      time4.google.com
    } & \makecell{
      diagnostics.meethue.com\\
      ecdinterface.philips.com
    } \\
    \hline
  \end{tabular}
  \caption{Domain names extracted by Mandalari \emph{et al.} \cite{blocking-without-breaking} (referred to as ``theirs'') and our work (referred to as ``ours''). Note that the domain names reported by Mandalari \emph{et al.} only concern non-essential hosts.}
  \label{tab:iotrimmer}
\end{table*}

A direct comparison of the domain names beyond the mere numbers is unfortunately not possible. This is because Mandalari \emph{et al.} were primarily interested in finding ``non-essential'' hosts, i.e. hosts that can be blocked, for example for privacy reasons, without negatively impacting the essential functionality of an IoT device. Consequently, while the total numbers are known, they only published the domain names of non-essential hosts, whereas we do not differ between essential and non-essential ones.
Nevertheless, we have also included the actual domain names in Table~\ref{tab:iotrimmer}. The differences between the two works can be explained by the fact that our approach actively looks for hidden communication patterns, the fact that their work is focused on identifying non-essential hosts, and the reasons already mentioned above. For example, the domain name \texturl{ecdinterface.philips.com} does not exist anymore (in 2025), while \texturl{diagnostics.meethue.com} could be part of a periodic or sporadic pattern filtered out by our approach. This can also explain the empty intersections between the two works for two of the tested devices.
%Besides two domains related to the TP-Link HS110,
%the sets of extracted domain names are disjoint.
%First, it must be noted that,
%for the list of their hosts,
%we could only consider the ones reported in their article,
%which are only the non-required ones.
%The required ones that they extracted might be part of those that we extracted.
%Then, two factors can explain this difference
%The first, simple one is the timing difference between both works:
%since their analysis, the manufacturers might have updated their devices,
%and changed the contacted endpoints.
%Furthermore, as per the three related works we compared to earlier,
%their approach is by design limited to exploring the first level of the signature tree.
%Our more thorough approach is thus able to gather additional information,
%as shown by the sheer number of extracted domains.

%\input{sections/discussion.tex}
\section{Related Work}
\label{sec:related-work}

Research on IoT device profiling has been fruitful in the last years.
Such works lie at differing layers of abstraction,
which can be classified along two orthogonal axes:
\begin{itemize}
    \item \emph{Object abstraction}: which semantic unit is profiled, from individual events,
    to single devices, to the IoT network as a whole.
    \item \emph{Behavior abstraction}: whether the work intends to profile the low-level network traffic,
    which can be further dissected into individual packets or aggregate flows,
    or the higher-level abstract device events retrieved from home automation systems.
\end{itemize}

% Following this classification,
% our work intends to profile \emph{individual events},
% at the \emph{network flow} level enhanced with application-layer data.

In the following,
we will summarize related works,
and position them along the aforementioned abstraction axes;
Table~\ref{tab:abstraction} shows the resulting classification.

\begin{table}
    \centering
    \begin{tabular}{|c|c|c|c|}
        % \cline{2-4}
        % \multicolumn{1}{c|}{} & \multicolumn{3}{c|}{\textbf{Object}} \\
        \cline{2-4}
        \multicolumn{1}{c|}{} & \textbf{Event} & \textbf{Device} & \textbf{Network} \\
        \hline
        \textbf{Packets} & \makecell{\emph{PingPong} \cite{ping-pong},\\\emph{D-interact} \cite{sun_inferring_2022}} & \emph{IoTa} \cite{duan_iota_2023} & Previous work \cite{smart-home-firewall} \\
        \hline
        \textbf{Flows}  & \makecell{Mandalari \cite{blocking-without-breaking},\\\textbf{This work}} & \makecell{MUD \cite{mud},\\Hamza \cite{hamza_clear_2018}} & \makecell{Previous work \cite{smart-home-firewall},\\Girish \cite{in-the-room}} \\
        \hline
        \textbf{Events} & / & \multicolumn{2}{c|}{\emph{BehavIoT} \cite{behaviot}} \\
        \hline
    \end{tabular}
    \caption{Abstraction level of device behavior models in our work and in the literature}
    \label{tab:abstraction}
\end{table}

Girish \textit{et al.} \cite{in-the-room} conducted the first comprehensive analysis
of LAN Smart Home traffic.
They exhaustively characterized the protocols used,
highlighting the security and privacy vulnerabilities,
of intra-network device communication.

Among the numerous proposals for IoT device behavior models,
the most widespread is probably the IETF's MUD standard \cite{mud},
as it has been standardized and backed up by multiple big players,
including Cisco \cite{mud-cisco}.
Subsequently, research leveraging this standard was fueled,
including Hamza \textit{et al.}'s solution to
generate MUD profiles from network traces of IoT devices' traffic \cite{hamza_clear_2018}.
The generated profiles, however, suffer from the shortcomings inherent to the MUD standard,
namely, among others, the lack of support for protocols other than
IP (v4 \& v6) on layer 3,
and TCP, UDP and ICMP on layer 4.
In previous work \cite{smart-home-firewall} proposed a syntax
to express the intended network behavior of devices,
inspired by MUD,
while overcoming some of its shortcomings,
and including support for traffic related to cross-device interactions.

Regarding research efforts to extract event signatures from network traffic,
three works follow an interact \& measure workflow similar to ours:
\emph{PingPong} \cite{ping-pong},
\emph{D-interact} \cite{sun_inferring_2022},
and \emph{IoTa} \cite{duan_iota_2023}.
Nevertheless, all three lie at a finer granularity level than our work,
as their signatures are composed of individual packets,
and leverage packet-level metadata,
such as the packet size.
We argue that such features can be subject to network churn
and therefore vary from one event execution to the other,
making them imprecise to accurately characterize network behavior.

Mandalari \textit{et al.} \cite{blocking-without-breaking}
centered their analysis on the hosts contacted by Smart Home devices,
characterized mainly by their domain names,
but falling back to their plain IP address if the domain name was unavailable.
Their goal was to identify non-required hosts,
i.e. hosts which could be blocked without impairing the device's operation,
e.g. telemetry or advertisement services.
While they designed a workflow similar to ours to identify the successful events,
they did not further explore the event signature tree.

Hu \textit{et al.} proposed \emph{BehavIoT} \cite{behaviot},
a system which models the behavior of a whole Smart Home network,
with the intent of using it to detect when the network produces unintended behavior,
potentially due to unwanted communication, or even a security breach.
Their model is twofold:
on the one hand, they model individual devices events;
on the other hand, cross-device interactions.
Their complete model is generated based on the network traffic produced by the devices,
and takes the form of a probabilistic Finite State Machine (FSM),
representing all the possible event transitions in the home network.

%By covering all individual device events as well as cross-device interactions,
%their work exhibit a broad scope;
%yet, the model is \emph{ad hoc},
%as it can only represent the network in which it was generated.

All the above-mentioned works have in common that, unlike our work, they do not actively attempt to discover hidden communication patterns.

% \subsection{Model-based Smart Home security systems}

% Various research works have leveraged Smart Home network models
% similar to the aforementioned ones
% as a component of a security system,
% intended to protect the home network from unintended traffic.

% Duan \textit{et al.}, for their \emph{IoTa} system \cite{duan_iota_2023},
% apply a strategy similar to \emph{PingPong}:
% they extract packet-level signatures of devices,
% only considering their length and direction.
% Subsequently, they model the complete device's network pattern with an FSM,
% each node representing one packet signature.
% Additionally, they leveraged the models
% as a traffic monitoring system:
% packets which do not comply with the FSM
% are flagged as unwanted.

% De Keersmaeker \textit{et al.} \cite{smart-home-firewall}
% defined a syntax to express to intended behavior of devices,
% inspired by MUD,
% while overcoming some of MUD's shortcomings,
% and including support for traffic related to cross-device interactions
% They then built a deny-list firewall system,
% taking their profiles as configuration files,
% and leveraging the NFTables Linux firewall \cite{nftables};
% we used their firewall as a building block of our framework.

%%% New subsection about Chaos Engineering (from rebuttal)

%\subsection*{\new{Relation with Chaos Engineering}}

Finally, the concept of chaos engineering \cite{chaos-engineering} is related to our approach.
It consists in injecting faults (e.g. network loss or latency)
or heavy load (e.g. on CPU, memory, etc.).
% Examples of tools for chaos engineering include Chaos Monkey \cite{chaosmonkey},
% Chaos Mesh \cite{chaosmesh},
% and Steadybit \cite{steadybit}.
While our approach shares some concepts with chaos engineering,
its novelty resides in the fact that we are the first to apply it
to this extent in the context of smart homes.
Moreover, today, no chaos engineering tool provides control over network traffic at a protocol level as precise as us,
by blocking traffic based on specific fields.
The most precise is Steadybit \cite{steadybit},
which allows blocking TCP altogether.

\section{Conclusion}
\label{sec:conclusion}

In this article,
we presented a novel methodology and framework
to extract comprehensive, multi-level traffic signatures
from Smart Home device events, whereas existing works are limited to first-level signatures. The signatures take the form of trees, with the nodes being the constitutive traffic flows.
By the usage of a dynamic blocking scheme, our approach allows uncovering \emph{hidden} traffic patterns that only occur when the default communication mechanisms of a device fail.
%tampering with the network the devices are connected to. 

We show that more than half of the tested devices
exhibit such hidden traffic patterns. 
In average, two hidden flows are discovered for each device, across all 
studied interactions with the device. 
Our work highlights the robustness capabilities of Smart Home devices,
by proposing a new metric dubbed the \emph{robustness score}.
We can envision that,
in the future,
such a score might be used as a marketing argument;
indeed, customers would prefer a device which is able to function
even in inhospitable network conditions.

% The implications for Smart Home network security solutions are primordial.
% Indeed, whereas they usually only consider first-level signatures,
% they should encompass the whole tree depth to provide an exhaustive protection,
% the latter being paramount in networks so close to the user's personal space.
% We hope to fuel further research in this direction,
% either by covering the shortcomings of our solution,
% or by applying a similar methodology to other scenarios,
% with the global incentive of making our networks more secure.

As future work, we would like to apply our methodology to scenarios where multiple Smart Home devices are involved in a single interaction.  This is conceptually supported by our framework, but will require an extension of the mechanisms we use for event triggering. In this case, approaches to further automate the latter should also be investigated.
Besides, our robustness analysis only focused on binary traffic verdict,
i.e. either the traffic passes or is blocked.
To further extend our robustness analysis,
we could consider other traffic disturbances,
such as limiting the traffic rate and inserting traffic bursts.
A study covering such bursts would be transversal to ours,
focusing more on robustness with respect to, e.g., DoS attacks.

% Acknowledgments
\section*{Acknowledgments}

F. De Keersmaeker is an F.R.S.-FNRS Research Fellow.
R. Sadre is supported by the Walloon Belgian region project CyberExcellence.
C. Pelsser is supported by the Cisco University Research Program Fund,
an advised fund of Silicon Valley Foundation,
and the Win4Collective CARAPACE project.
All icons are from \rurl{flaticon.com}.

\bibliographystyle{IEEEtran}
\bibliography{biblio}

\appendix
\subsection{Statement on ethical issues}
No ethical issues were raised by this work.
Our device profiling activities are in line with the EU legislation on reverse engineering
\cite{eu-law-reverse,eu-law-interpretation}.
All experiments were performed in a controlled environment,
and did not involve any user personal data or other privacy concerns.

\subsection{List of protocol fields supported by the firewall}

Table~\ref{tab:protocol-fields} displays all the protocol fields which can be leveraged by the firewall to block packets.

\begin{table}
  \centering
  \begin{tabular}{c|c|c|c}
    \textbf{Layer} & \textbf{Protocol} & \textbf{Field} & \textbf{Description} \\
    \hline
    \multirow{2}{*}{Net.} & IPv4, & src, & Source \& destination host \\
     & IPv6 & dst & (IP address or domain name) \\
    \hline
    \multirow{2}{*}{Trans.} & TCP, & src-port, & Source \& destination port \\
     & UDP & dst-port & (linked to their respective host) \\
    \hline
    \multirow{8}{*}{App.} & \multirow{2}{*}{(m)DNS} & qtype & Query type (A, CNAME, etc.) \\
     & & qname & Queried domain name \\
     \cline{2-4}
     & \multirow{3}{*}{HTTP} & is\_response & Is it an HTTP response ? \\
     & & method & HTTP method (GET, PUT, etc.) \\
     & & uri & Requested URI \\
     \cline{2-4}
     & \multirow{3}{*}{CoAP} & is\_response & Is it a CoAP response ? \\
     & & code & CoAP code (GET, PUT, etc.) \\
     & & uri\_path & Requested URI \\
  \end{tabular}
  \caption{List of protocol fields, per layer, which can be leveraged by the firewall for its verdict}
  \label{tab:protocol-fields}
\end{table}

\subsection{Event signature tree size without pruning}
\label{app:no-pruning}

In this appendix,
we give additional insight advocating for our event signature tree pruning heuristic.
We generated a sample tree,
by performing preliminary experiments with the \emph{toggle} interaction
of the TP-Link HS110 \cite{hs110}.
The size of the complete tree was huge:
it comprised 75 Flow IDs in total.
However, among these, only five were unique.
As explained in the main text (Section~\ref{sec:tree-pruning}),
each occurrence of the same node had the same set of children Flow IDs.
We deem our \emph{node pruning} heuristic will not lose information,
and is efficient in reducing the tree to a manageable size.

\subsection{Additional results}
\label{app:full_results}

Fig.~\ref{fig:count_flow_id} shows the count of unique Flow IDs discovered by our framework
for each investigated event,
split into \emph{first-level} and \emph{hidden} Flow IDs.
%Fig.~\ref{fig:robustness-score} shows the \emph{robustness score} for each event.
% Fig.~\ref{fig:count_dns_all} depicts the count of
% domain names and DNS resolvers contacted by each event,
% both split into \emph{first-level} and \emph{hidden}.
The x-axis pertains to the event ID,
as provided in Table~\ref{tab:devices_id}.

\begin{figure}
  \centering
  \includegraphics{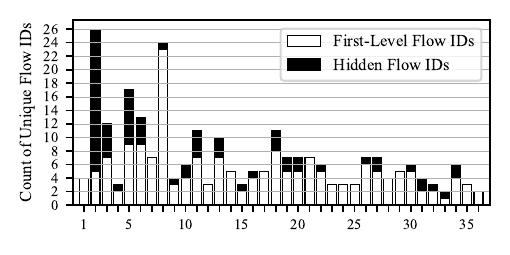}
  \caption{
    Unique Flow IDs.
    The x-axis shows the \emph{Event ID}.
  }
  \label{fig:count_flow_id}
\end{figure}

% \begin{figure}
%   \centering
%   \includegraphics{figures/graphs/count_dns_all.pdf}
%   \caption{Count of unique domain names / DNS resolvers contacted across all events.}
%   \label{fig:count_dns_all}
% \end{figure}

\end{document}